# Ab-initio insights into the physical properties of $X$Ir$_3$ ($X$ = La, Th) superconductors: A comparative analysis

Md. Sajidul Islam[1,2], Razu Ahmed[1,2], M.M. Hossain[2,3], M.A. Ali[2,3], M.M. Uddin[2,3], S.H. Naqib[1,2*]

[1]Department of Physics, University of Rajshahi, Rajshahi 6205, Bangladesh

[2]Advanced Computational Materials Research Laboratory, Department of Physics, Chittagong University of Engineering and Technology (CUET), Chattogram-4349, Bangladesh

[3]Department of Physics, Chittagong University of Engineering and Technology (CUET), Chattogram-4349, Bangladesh

*Corresponding author, Email: salehnaqib@yahoo.com

**Abstract**

Here we report the structural, elastic, bonding, thermo-mechanical, optoelectronic and superconducting state properties of recently discovered $X$Ir$_3$ ($X$ = La, Th) superconductors utilizing the density functional theory (DFT). The elastic, bonding, thermal and optical properties of these compounds are investigated for the first time. The calculated lattice and superconducting state parameters are in reasonable agreement to those found in the literature. In the ground state, both the compounds are mechanically stable and possess highly ductile character, high machinability, low Debye temperature, low bond hardness and significantly high melting point. The thermal conductivities of the compounds are found to be very low which suggests that they can be used for thermal insulation purpose. The population analysis and charge density distribution map confirm the presence of both ionic and covalent bonds in the compounds with ionic bond playing dominant roles. The calculated band structure and DOS profiles indicate metallic character of $X$Ir$_3$ compounds. It is also found from the DOS profile that the Ir 5$d$-states dominate near the Fermi level. Unlike the significant anisotropy observed in elastic and thermal properties, all the optical constants of $X$Ir$_3$ compounds exhibit almost isotropic behavior. The optical parameters' profiles correspond very well with the electronic band structure and DOS features. We have estimated the superconducting transition temperature ($T_c$) of $X$Ir$_3$ compounds in this work. The calculated values of $T_c$ for LaIr$_3$ and ThIr$_3$ compounds are 4.91 K and 5.01 K, respectively. We have compared and contrasted all the physical properties of $X$Ir$_3$ ($X$ = La, Th) compounds in this study.

## 1. Introduction

Superconductivity is one of the most interesting and significant topics in both theoretical and experimental condensed matter physics. Nowadays the investigation of superconductors has piqued the interest of modern society owing to their features suitable for wide range of applications. Many binary systems made of rare earth and transition metals have been found to exhibit superconductivity. The addition of rare earth and 5$d$ transition metal element such as Ir (Iridium) can create a compound that displays superconductivity and other attractive electronic, mechanical and optical features [1 - 4]. LaIr$_3$, ThIr$_3$, CeIr$_3$, CaIr$_2$, IrGe, and Mg$_{10}$Ir$_{19}$B$_{16}$ are a few examples of such compounds, where the superconductivity arises mainly from the Ir-5$d$ electronic states close to the Fermi level [1,4–8].

In this paper, the physical properties of two superconducting compounds $X$Ir$_3$ ($X$ = La, Th) [1,4], that crystallizes in a rhombohedral structure with space group *R-3m*, have been studied. In this work, we investigate the structural, elastic, bonding, thermo-mechanical, optoelectronic and superconducting state properties of these two promising compounds. The obtained results were compared to existing experimental and theoretical data, where applicable. It is worth noting that many technologically essential physical features of $X$Ir$_3$ ($X$ = La, Th) materials are still unknown in comparison to other systems such as chalcogenides, oxides, and pnictides. We focus particularly on the unexplored physical characteristics of $X$Ir$_3$ ($X$ = La, Th).

Recently, the superconducting compound LaIr$_3$ was synthesized and characterized experimentally by Bhattacharyya et al. [9]. They investigated the superconducting ground state in LaIr$_3$ and observed that it is a bulk type-II superconductor with $T_C$ = 2.5 K. Furthermore, Haldolaarachchige et al. [1] examined the physical characteristics of LaIr$_3$ both experimentally and theoretically. They suggested that LaIr$_3$ is a BCS-type superconductor which exhibits superconductivity at 3.3 K. Moreover, they discovered that the superconductivity in LaIr$_3$ is arising from the Ir-5$d$ states at the Fermi surface. Górnicka et al. [4] prepared a polycrystalline sample of ThIr$_3$ and examined its electronic and magnetic characteristics using dc magnetization, electrical resistivity, and heat capacity assessment. They concluded that ThIr$_3$ is a $d$-band, moderately coupled type-II superconductor with $T_c$ = 4.41 K. They also stated that the 5$d$-states of Ir play the most important role in the appearance of superconductivity in ThIr$_3$ with some contributions from the Th-6$d$ states. Moreover, Górnicka et al. [5] also synthesized and investigated the superconducting state properties of CeIr$_3$. Based on the specific-heat data analysis, they estimated that CeIr$_3$ is a moderately coupled type-II superconductor which is superconducting below 2.5 K. There are few other related studies on the materials CaIr$_2$ [6], IrGe [7], and Mg$_{10}$Ir$_{19}$B$_{16}$ [8] where the superconductivity is derived from the Ir-$d$ electronic orbitals.

Several physical characteristics, such as structural characterization, band structure calculation, electronic energy density of states, Fermi surface, and superconducting state features of $X$Ir$_3$ ($X$ = La, Th) have been investigated theoretically as well as experimentally so far [1,4,9]. To the best of our knowledge, most of the physical features of these compounds, such as elastic, chemical bonding, charge density distribution, Mulliken population analysis, thermo-mechanical properties, and energy dependent optical constants have not been studied so far. For instance, no citable work has yet to be found on the elastic characteristics of these compounds such as the Poisson's ratio, machinability index, Kleinman parameter, hardness, anisotropy in elastic moduli etc. Moreover, the various thermo-mechanical parameters of these compounds such as the Debye temperature, lattice thermal conductivity, melting temperature, sound velocities, and thermal expansion coefficient have not been explored at all. In addition, still there are lack of theoretical understanding of the bonding and optical characteristics of these materials.

Therefore, it is obvious from the preceding discussion that many physical features of these compounds remain unexplored. In this study, we want to overcome this research gap, which is the key motivation for the current investigations. Hence, it is our crucial theoretical interest to study the structural, elastic, mechanical, bonding, acoustic, optoelectronic, thermal and superconducting state properties of $X$Ir$_3$ ($X$ = La, Th) compounds thoroughly in this paper. It is worth keeping in mind that information on all these hitherto unexplored physical properties are essential to unlock the potential of $X$Ir$_3$ ($X$ = La, Th) for possible applications.

The remaining part of this study is organized as follows: In Section 2, we provide an overview of our computational scheme. We discussed the calculated results of various properties of $X$Ir$_3$ ($X$ = La, Th) compounds in Section 3. Finally, the major findings of this investigation are summarized in Section 4.

## 2. Computational scheme

The CASTEP (CAmbridge Serial Total Energy Package) code [10] is a comprehensive software package which is designed to calculate the ground state energy of a system as well as various other physical properties based on the quantum mechanical density functional theory (DFT) [11]. The ground state of a crystalline system can be obtained in DFT formalism by solving the Kohn-Sham equation [12]. In this investigation, we employed the local density approximation (LDA) [13] for the exchange-correlation energy calculations because it yields better outcomes of ground state structural properties for $X$Ir$_3$ ($X$ = La, Th) compounds than the other schemes including the Perdew-Burke-Ernzerhof generalized gradient approximation (PBE-GGA) [14]. To model the coulomb potential energy that arises from the interaction of valence electrons with ion cores of $X$ ($X$ = La, Th) and Ir atoms, we employed Vanderbilt-type ultra-soft pseudopotential [15]. Ultra-soft pseudopotential saves a large amount of processing time while compromising just a little bit of computational accuracy. To find the lowest energy structure, we adopted The BFGS (Broyden-Fletcher-Goldferb-Shanno) minimization technique [16] for geometry optimization.

To execute the pseudo atomic operations, the valence electron configurations used in this study were [$5s^2 5p^6 5d^1 6s^2$] for the La atom, [$6s^2 6p^6 6d^2 7s^2$] for the Th atom, and [$5d^7 6s^2$] for the Ir atom, respectively. In this work, the plane-wave cut off energy was set at 400 eV to obtain the total energy convergence. The sampling of the Brillouin zone of $X$Ir$_3$ compounds was performed using a mesh size of 11×11×2 k-points based on the Monkhorst-Pack scheme [17]. Employing finite basis set corrections [18], the geometry optimization of LaIr$_3$ was accomplished with convergence thresholds of $0.5 \times 10^{-5}$ eV/atom for the total energy, $0.5 \times 10^{-3}$ Å for maximum ionic displacement, 0.02 GPa for maximum stress component and 0.01 eV/Å for maximum ionic force. On the other hand, for the ThIr$_3$ compound, the geometry optimization was performed with convergence within $10^{-5}$ eV/atom for the total energy, within $10^{-3}$ Å for maximum ionic displacement, within 0.05 GPa for maximum stress component and within 0.03 eV/Å for maximum ionic force. These chosen tolerance levels provide very good results of structural, elastic and optoelectronic properties with an optimum computing speed. In this work, all the DFT computations of $X$Ir$_3$ ($X$ = La, Th) compounds were carried out in the ground state with default temperature (0 K) and pressure (0 GPa).

After the geometry optimization, the elastic constants, Mulliken bond population, charge density distribution, band structure, density of states (DOS) and optical properties were evaluated. The stress-strain approach [19] was used to compute the independent elastic constants, $C_{ij}$, of rhombohedral crystal structure. The computations provide six independent elastic constants. The bulk modulus $B$, shear modulus $G$, and many other elastic properties were calculated from these independent elastic constants utilizing the Voigt-Reuss-Hill (VRH) method [20,21]. The computed elastic constants are used to calculate various thermophysical parameters such as Debye temperature, thermal conductivity, melting temperature, thermal expansion coefficient, heat capacity, and so on. The Mulliken bond population analysis was used in order to gain a better knowledge of the bonding features of $XIr_3$ ($X$ = La, Th) compounds using a projection of the plane-wave states onto a linear combination of atomic orbital basis sets [22–24]. The Mulliken bond population analysis was accomplished with the Mulliken density operator, which is written on an atomic (or quasi-atomic) basis in the following manner:

$$P_{\mu\nu}^{M}(g) = \sum_{g'} \sum_{\nu'} P_{\mu\nu'}(g') S_{\nu'\nu}(g - g') = L^{-1} \sum_{k} e^{-ikg} (P_k S_k)_{\mu\nu'} \tag{1}$$

and the total charge that exists on an atomic species $A$ can be stated as:

$$Q_A = Z_A - \sum_{\mu \in A} P_{\mu\mu}^{m}(o) \tag{2}$$

where $Z_A$ refers to the charge of the atomic core.

All the optical parameters of $XIr_3$ ($X$ = La, Th) in the ground state can be fully investigated by utilizing the complex dielectric function, $\varepsilon(\omega) = \varepsilon_1(\omega) + i\varepsilon_2(\omega)$. Utilizing the Kramers-Kronig equations, we can simply determine the real part of the dielectric function from its imaginary part. CASTEP computes the imaginary part, $\varepsilon_2(\omega)$, by employing the following formula [25,26]:

$$\varepsilon_2(\omega) = \frac{2e^2\pi}{\Omega \varepsilon_0} \sum_{k,\nu,c} |\langle \psi_k^c | \hat{u}\cdot\vec{r} | \psi_k^\nu \rangle|^2 \delta(E_k^c - E_k^\nu - E) \tag{3}$$

In the above equation, $\omega$ refers to the angular frequency of the electromagnetic wave, $\Omega$ denotes the volume of the unit cell, $e$ is the electronic charge, and at a particular wave-vector $k$, $\psi_k^\nu$ and $\psi_k^c$ represent the valence and conduction band wave functions, respectively. The other optical parameters, such as the refractive index, reflectivity, absorption coefficient, optical conductivity, and energy loss function can be evaluated from the complex dielectric function, $\varepsilon(\omega)$ by employing the following set of relations [27,28]:

$$n(\omega) = \frac{1}{\sqrt{2}} [\{\varepsilon_1(\omega)^2 + \varepsilon_2(\omega)^2\}^{1/2} + \varepsilon_1(\omega)]^{1/2} \tag{4}$$

$$k(\omega) = \frac{1}{\sqrt{2}} [\{\varepsilon_1(\omega)^2 + \varepsilon_2(\omega)^2\}^{1/2} - \varepsilon_1(\omega)]^{1/2} \tag{5}$$

$$R(\omega) = \left|\frac{\tilde{n}-1}{\tilde{n}+1}\right| = \frac{(n-1)^2 + k^2}{(n+1)^2 + k^2} \tag{6}$$

$$\alpha(\omega) = \frac{4\pi k(\omega)}{\lambda} \tag{7}$$

$$\sigma(\omega) = \frac{2W_{cv}\hbar\omega}{\vec{E_0}^2} \tag{8}$$

$$L(\omega) = Im(-\frac{1}{\varepsilon(\omega)}) \tag{9}$$

where, $n(\omega)$ and $k(\omega)$ are the real and imaginary parts of the complex refractive index, $R(\omega)$ is the reflectivity, $\alpha(\omega)$ is the absorption coefficient, $\sigma(\omega)$ is the optical conductivity, $W_{cv}$ refers to the transition probability per unit time and $L(\omega)$ is the energy loss function.

## 3. Results and discussions

### *3.1. Structural properties*

As we described in the earlier that $X$Ir$_3$ ($X$ = La, Th) crystallize in the rhombohedral structure (space group *R-3m*, No.166) [1,4]. The schematic view of the optimized unit cell of our chosen compounds is depicted in Figure 1. This unit cell contains 36 atoms (9 *X* and 27 Ir) and has nine formula units (Z = 9). It consists of three unique positions of Ir atoms and two unique positions of *X* (*X* = La, Th) atoms. Atomic environments of all the sites in $X$Ir$_3$ compounds are Frank-Kasper polyhedron [29]. In one of its positions, *X*-site is surrounded by sixteen Ir atoms, while in the other *X*-site is centered by twelve Ir atoms.

The geometry optimization yields the fully relaxed lattice parameters of the chosen compounds. Table 1 summarized these optimized lattice parameters together with available experimental findings for comparison. It is seen from the Table 1 that the values of *a*, *c* and $V_0$ calculated by the functional LDA deviate from the experimental values by not more than 0.86%, 0.35% and 2.05%, respectively. These findings suggest that the experimental data and the theoretical outcomes are in fair agreement. On the other hand, the values of *a*, *c* and $V_0$ calculated by the functional GGA(PBE) differ from the reported experimental values by more than 1.23%, 1.82% and 4.35%, respectively. For this reason, we estimated the elastic, mechanical, bonding, optoelectronic and thermal properties of $X$Ir$_3$ compounds using the LDA approach.

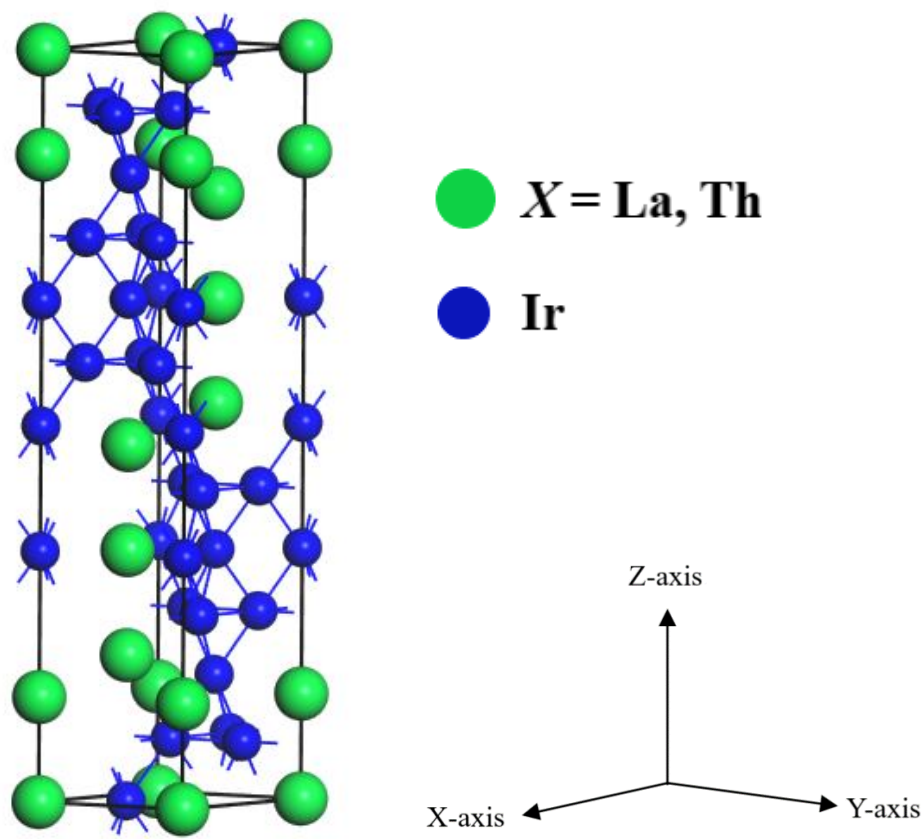

**Figure 1.** Three-dimensional schematic illustration of the conventional unit cell of $X$Ir$_3$ ($X$ = La, Th) compounds.

**Table 1.** The optimized unit cell parameters ($a = b$ and $c$ in Å, optimized cell volume $V_0$ in Å$^3$) of $X$Ir$_3$ ($X$ = La, Th) compounds along with corresponding experimental (Expt.) values.

| Compound | Functionals | $a = b$ | $c$ | $c/a$ | $V_0$ | Ref. |
|---|---|---|---|---|---|---|
| LaIr$_3$ | LDA | 5.355 | 26.353 | 4.921 | 654.472 | This work |
| | GGA(PBE) | 5.458 | 26.881 | 4.925 | 693.499 | This work |
| | - | 5.320 | 26.340 | 4.951 | 645.609 | [1,4]$^{Expt.}$ |
| ThIr$_3$ | LDA | 5.294 | 26.330 | 4.974 | 638.997 | This work |
| | GGA(PBE) | 5.405 | 26.904 | 4.978 | 680.735 | This work |
| | - | 5.339 | 26.423 | 4.950 | 652.371 | [4]$^{Expt.}$ |

*3.2. Elastic and mechanical properties*

Elastic constants are the physical properties that measure the response of a material to external stress. They determine the stiffness of material, and are used to calculate how the material behaves when subjected to a given force on different crystal planes. A compound with rhombohedral crystal structure has six independent elastic constants ($C_{ij}$), which are $C_{11}$, $C_{12}$, $C_{13}$, $C_{14}$, $C_{33}$ and $C_{44}$. These single crystal elastic constants are utilized to calculate the various other elastic parameters of our chosen compounds. The computed values of the single crystal elastic constants of our chosen compounds are shown in Table 2. In order to be mechanically stable, a rhombohedral crystal structure must fulfill the following Born-Huang [30] stability criteria:

$C_{44} > 0;$

$(C_{11} - C_{12}) > 0;$

$(C_{11} - C_{12})C_{44} - 2C_{14}^2 > 0;$

$(C_{11} + C_{12})C_{33} - 2C_{13}^2 > 0 \qquad (10)$

At zero pressure, all of the computed elastic constants of $X\text{Ir}_3$ ($X$ = La, Th) compounds meet the above mechanical stability conditions. This implies that all of the compounds under study are mechanically stable.

**Table 2.** Computed elastic constants, $C_{ij}$ (in GPa) and Cauchy pressure ($C_{12} - C_{44}$), (in GPa) of $X\text{Ir}_3$ ($X$ = La, Th) compounds.

| Compound | $C_{11}$ | $C_{12}$ | $C_{13}$ | $C_{14}$ | $C_{33}$ | $C_{44}$ | ($C_{12} - C_{44}$) |
|---|---|---|---|---|---|---|---|
| LaIr$_3$ | 329.33 | 204.30 | 149.57 | 42.62 | 409.47 | 54.49 | 149.81 |
| ThIr$_3$ | 424.35 | 230.21 | 171.47 | 26.30 | 457.47 | 26.75 | 203.46 |

The results presented in Table 2 are novel and no comparison can be made. The independent elastic constants $C_{11}$ and $C_{33}$ determine the crystalline solid's responses to uniaxial strain whereas the resistance to shape deformation is described by the elastic constant $C_{44}$. The three off-diagonal elements, $C_{12}$, $C_{13}$ and $C_{14}$ describe the shape deforming shearing strain brought on by forces on various crystal planes. Table 2 shows that the elastic constants $C_{11}$ and $C_{33}$ for both the compounds are significantly greater than the elastic constant $C_{44}$. These findings suggest that shear deformation of the compounds is significantly easier than unidirectional strain along any of the three crystallographic axes. The compound ThIr$_3$ is stiffer than LaIr$_3$ when uniaxial elastic strengths are considered. The elastic constants $C_{11}$ and $C_{33}$ assess the crystal's ability to endure the uniaxial stress that imposed to the crystallographic $a(b)$- and $c$-directions, respectively. Table 2 shows that the value of $C_{33}$ is larger than $C_{11}$ for both the studied materials. This result is interesting since the *c*-axis lattice parameter of the unit cell is much larger than the *a(b)*-axis. We think that the cage-like structural units of Ir atoms along the *c*-direction gives the high mechanical strength of the $X$Ir$_3$ ($X$ = La, Th) compounds in this particular crystallographic direction. These results imply that the compounds' structures are more contractible along the *a(b)*-direction compared to the *c*-direction. Thus, it can be said that the atomic bonding in the *ab*-plane will be weaker compared to the out-of-plane directions. The large discrepancy in tensor $C_{11}$ and $C_{33}$ suggests that both the compounds have significant amount of elastic anisotropy. This discrepancy is much higher for the compound LaIr$_3$ compared to ThIr$_3$ which further implies that LaIr$_3$ is comparatively more anisotropic in nature. The difference between the elastic constants $C_{12}$ and $C_{44}$ is defined as the Cauchy pressure ($C_{12}$ - $C_{44}$), which is widely used to evaluate the type of chemical bonding in a crystal and to predict the ductile or brittle nature of solids [31]. It is observed from the Table 2 that both the compounds have large positive Cauchy pressure. This suggests that their chemical bonding is dominated by ionic/metallic bondings and they are expected to show high level of ductility.

The single crystal elastic constants may be used to calculate different polycrystalline elastic moduli, Pugh's ratio, and Poisson's ratio. To evaluate the values of polycrystalline bulk modulus (*B*), shear modulus (*G*), Young's modulus (*Y*), and Poisson's ratio (*η*), following equations [32] are used and the calculated results are listed in Table 3.

$$B = \frac{B_V + B_R}{2} \tag{11}$$

$$G = \frac{G_V + G_R}{2} \tag{12}$$

$$Y = \frac{9BG}{3B + G} \tag{13}$$

$$\eta = \frac{3B - 2G}{2(3B + G)} \tag{14}$$

The Voigt [33] and Reuss [34] values of the bulk modulus and shear modulus are represented in the above equations by the symbols $B_V$, $B_R$ and $G_V$, $G_R$, respectively.

The bulk and shear moduli of crystals can be utilized to evaluate their overall mechanical behaviors. Bulk modulus describes the resistance of a material to volume deformation by hydrostatic pressure while shear modulus measures the resistance to shape deformations caused by shearing stress [35]. The larger value of bulk modulus (*B*) compared to shear modulus (*G*) (see Table 3), suggesting that the mechanical strengths of $X$Ir$_3$ compounds will be governed by the shearing strain.

The Young's modulus of a material can be used to estimate its stiffness and thermal shock resistance [36,37]. From Table 3, it is seen that the compound LaIr$_3$ has a lower value of *Y* than the compound ThIr$_3$ indicating that ThIr$_3$ is stiffer than LaIr$_3$. Since the critical thermal shock resistance is inversely proportional to the Young's modulus [38] it can be suggested that among the two compounds LaIr$_3$ will exhibit better thermal shock resistance.

The failure mode (ductile or brittle nature) of a given system can be explained from the ratio of *B* to *G*, which is called the Pugh's ratio (*B/G*) [39]. Pugh predicted that if this ratio is less than 1.75, a material will be brittle; otherwise, the material will be ductile. From Table 3, it is noticed that the *B/G* ratio of our two studied compounds is much higher than 1.75, which suggests once again that they are highly ductile in nature.

The Poisson's ratio (*η*) is another elastic parameter that can be used to understand nature of the bonding forces and to predict the crystal stability against shear. Materials with low values of Poisson's ratio possess stability against shear [40]. Thus, all our chosen compounds should be unstable against shear because they have high Poisson's ratio. Again, it is known that the stability of the crystal systems is governed by the central forces if the Poisson's ratio lies between 0.25 and 0.50 [41]. Therefore, from Table 3 it can be said that the interatomic forces in $X$Ir$_3$ ($X$ = La, Th) compounds are the central forces. Moreover, the Poisson's ratio can predict the ductility or brittle nature of a crystal as well as the type of chemical bonding in a crystal. Brittle failure occurs in solids when *η* is less than a critical value of 0.26, while ductile failure occurs with *η* greater than 0.26 [38,42]. Thus, from Table 3, it can be concluded that studied compounds should show ductility. This result confirms the prediction from the analysis of Cauchy pressure and Pugh's ratio. A crystal with a Poisson's ratio of 0.1 is considered as purely covalent crystal, whereas one with a Poisson's ratio of 0.33 is considered as metallic compound [43]. Since both the compounds have η greater than 0.33 they are characterized by metallic bonding. Furthermore, the plasticity of a solid against shear can also be estimated from the Poisson's ratio. High value of η indicates that our compounds will have a high degree of plasticity.

**Table 3.** Calculated bulk modulus ($B_V$, $B_R$, *B* in GPa), shear modulus ($G_V$, $G_R$, *G* in GPa), Young's modulus (*Y* in GPa), Pugh's ratio (*B/G*), and Poisson's ratio (*η*) of $X$Ir$_3$ ($X$ = La, Th) compounds.

| Compound | $B_V$ | $B_R$ | $B$ | $G_V$ | $G_R$ | $G$ | $Y$ | $B/G$ | $\eta$ |
|---|---|---|---|---|---|---|---|---|---|
| LaIr$_3$ | 230.56 | 230.37 | 230.47 | 71.95 | 32.24 | 52.10 | 145.35 | 4.42 | 0.39 |
| ThIr$_3$ | 272.50 | 272.33 | 272.42 | 78.98 | 36.54 | 57.76 | 161.84 | 4.72 | 0.40 |

Table 3 confirms that the bulk mechanical strength of ThIr$_3$ is higher than that of LaIr$_3$.

In order to calculate the dimensionless Kleinman parameter ($\zeta$), machinability index ($\mu^M$), bulk modulus under uniaxial strain ($B_a$, $B_b$ and $B_c$) and isotropic bulk modulus ($B_{relax}$), the following formulae are employed [44,45] and the outcomes are given in Table 4.

$$\zeta = \frac{C_{11}+8C_{12}}{7C_{11}+2C_{12}} \tag{15}$$

$$\mu^M = \frac{B}{C_{44}} \tag{16}$$

$$B_a = a\frac{dP}{da} = \frac{\Lambda}{1+\alpha+\beta} \tag{17}$$

$$B_b = b\frac{dP}{db} = \frac{B_a}{\alpha} \tag{18}$$

$$B_c = c\frac{dP}{dc} = \frac{B_a}{\beta} \tag{19}$$

$$B_{relax} = \frac{\Lambda}{(1+\alpha+\beta)^2} \tag{20}$$

with, $\Lambda = C_{11}+2C_{12}\alpha+C_{22}\alpha^2+2C_{13}\beta+C_{33}\beta^2+2C_{33}\alpha\beta$

$$\alpha = \frac{\{(C_{11}-C_{12})(C_{33}-C_{13})\}-\{(C_{23}-C_{13})(C_{11}-C_{13})\}}{\{(C_{33}-C_{13})(C_{22}-C_{12})\}-\{(C_{13}-C_{23})(C_{12}-C_{23})\}}$$

$$\beta = \frac{\{(C_{22}-C_{12})(C_{11}-C_{13})\}-\{(C_{11}-C_{12})(C_{23}-C_{12})\}}{\{(C_{22}-C_{12})(C_{33}-C_{13})\}-\{(C_{12}-C_{23})(C_{13}-C_{23})\}}$$

**Table 4.** The dimensionless Kleinman parameter ($\zeta$), machinability index ($\mu^M$), isotropic bulk modulus ($B_{relax}$ in GPa), and bulk modulus under uniaxial strain ($B_a$, $B_b$ and $B_c$ in GPa of $X$Ir$_3$ ($X$ = La, Th).

| Compound | $\zeta$ | $\mu^M$ | $B_{relax}$ | $B_a$ | $B_b$ | $B_c$ |
|---|---|---|---|---|---|---|
| LaIr$_3$ | 0.72 | 4.23 | 285.99 | 829.38 | 829.38 | 921.53 |
| ThIr$_3$ | 0.66 | 10.18 | 337.63 | 1043.28 | 1043.28 | 957.14 |

A compound's resistance to stretching and bending forces can be estimated from the Kleinman parameter ($\zeta$) [32,44]. Its value typically lies between 0 and 1. It is seen from the Table 4 that both the compounds have $\zeta$ value higher than 0.5. This result inferred that mechanical strength in $X$Ir$_3$ ($X$ = La, Th) compounds is mainly provided by the bond bending contributions. The machinability index ($\mu^M$) is another parameter with which we can estimate how easily a material can be machined. This index can be used to compare the machinability of different materials and to determine the optimal cutting parameters for a given material. Furthermore, this index can also be used to assess the plasticity and dry lubricating feature of a material [32]. Table 4 shows that both the compounds have very high machinability index. The machinability index ThIr$_3$ is in fact extraordinarily high. Both the compounds are expected to possess excellent dry lubricating properties, lower friction value and lower feed forces. Moreover, the levels of machinability of ThIr$_3$ and LaIr$_3$ are significantly higher than many technologically important

MAX and MAB phase compounds [46–49].

To understand a material's mechanical behavior in practical applications it is essential to calculate its hardness value. There are a number of theoretical schema for determining the hardness as a function of Poisson's ratio ($\eta$), Young's modulus ($Y$), bulk modulus ($B$), and shear modulus ($G$). Some of the prominent schemes are given by N. Miao et al. [50,51], X. Chen et al. [52], Y. Tian et al. [53], and D. M. Teter [54], and E. Mazhnik et al. [55]. Using these approaches, we have calculated the values of $(H_V)_{micro}$, $(H_V)_{macro}$, $(H_V)_{Tian}$, $(H_V)_{Teter}$, and $(H_V)_{Mazhnik}$ via Eqns. (21) – (25). The obtained values are disclosed in Table 5.

$$(H_V)_{micro} = \frac{(1-2\eta)Y}{6(1+\eta)} \tag{21}$$

$$(H_V)_{macro} = 2[\left(\frac{G}{B}\right)^2 G]^{0.585} - 3 \tag{22}$$

$$(H_V)_{Tian} = 0.92(G/B)^{1.137} G^{0.708} \tag{23}$$

$$(H_V)_{Teter} = 0.151 G \tag{24}$$

$$(H_V)_{Mazhnik} = \gamma_0 \chi(\eta) Y \tag{25}$$

where, $\chi(\eta)$ is a function of the Poisson's ratio ($\eta$), which can be obtained from:

$$\chi(\eta) = \frac{1 - 8.5\eta + 19.5\eta^2}{1 - 7.5\eta + 12.2\eta^2 + 19.6\eta^3}$$

and $\gamma_0$ is a dimensionless constant with a value of 0.096.

It is observed from the Table 5 that, the scheme suggested by X. Chen et al. yields the lowest values of hardness while the scheme suggested by Mazhnik et al. yields the highest values of hardness for both the compounds. It can also be estimated from the Table 5 that the compound ThIr$_3$ is fairly harder than the compound LaIr$_3$. This result is consistent with the results on elastic parameters.

The formation of cracks, particularly in metals and ceramic materials, is one of the major issues with surface hard coatings on heavy-duty devices. Fracture toughness, $K_{IC}$, of a material is a measure of its resistance to surface crack formation. This parameter plays a vital role in the design of engineering materials. The following equation is used to calculate the fracture toughness ($K_{IC}$) [56]:

$$K_{IC} = \alpha_0^{-1/2} V_0^{1/6} [\xi(\eta) Y]^{3/2} \tag{26}$$

where, $V_0$ is the atomic volume; $\alpha_0$ = 8840 GPa; $\xi(\eta)$ is a function of the Poisson's ratio ($\eta$) which can be found from:

$$\xi(\eta) = \frac{1-13.7\eta+48.6\eta^2}{1-15.2\eta+70.2\eta^2-81.5\eta^3}$$

The fracture toughness is also shown in Table 5. It is seen from the Table 5 that the value of $K_{IC}$ of ThIr$_3$ compound is higher than LaIr$_3$ compound. This result suggests that ThIr$_3$ will offer higher resistance to surface crack formation than LaIr$_3$ and ThIr$_3$. The mechanical hardness, on the other hand, of the two compounds is a fairly similar.

**Table 5.** Calculated hardness (in GPa) with different schemes and fracture toughness (in MPam$^{1/2}$) of $X$Ir$_3$ ($X$ = La, Th) compounds.

| Compound | $(H_V)_{micro}$ | $(H_V)_{macro}$ | $(H_V)_{Tian}$ | $(H_V)_{Teter}$ | $(H_V)_{Mazhnik}$ | $K_{IC}$ |
|---|---|---|---|---|---|---|
| LaIr$_3$ | 3.83 | 0.55 | 2.79 | 7.87 | 8.37 | 1.85 |
| ThIr$_3$ | 3.85 | 0.50 | 2.79 | 8.72 | 9.32 | 2.32 |

The anisotropy in elasticity of a crystalline solid describes the directional dependence of its mechanical characteristics. Elastic anisotropy influences many physical processes, including the microcrack formation in solids, the development of plastic deformations in crystals, crack propagation dynamics, etc. Understanding elastic anisotropy has important consequences in both crystal physics and applied engineering fields. Therefore, it is essential to calculate the elastic anisotropy factors of our studied compounds in detail. For a thorough knowledge of elastic anisotropy, we have determined the Zener anisotropy parameter ($A$), shear anisotropy parameters ($A_1$, $A_2$, and $A_3$), anisotropy in compressibility ($A_B$), anisotropy in shear ($A_G$), the universal anisotropy index ($A^U$, $d_E$), the equivalent Zener anisotropy index ($A^{eq}$), the universal log-Euclidean index ($A^L$) and the anisotropies of the bulk modulus along $a$- and $c$-axes relative to the $b$-axis ($A_{Ba}$ and $A_{Bc}$) by using the following equations [40,41,44] and the outcomes are disclosed in Table 6 and Table 7.

$$A = \frac{2C_{44}}{C_{11}-C_{12}} \tag{27}$$

$$A_1 = \frac{4C_{44}}{C_{11}+C_{33}-2C_{13}} \tag{28}$$

$$A_2 = \frac{4C_{55}}{C_{22}+C_{33}-2C_{23}} \tag{29}$$

$$A_3 = \frac{4C_{66}}{C_{11}+C_{22}-2C_{12}} \tag{30}$$

$$A_B = \frac{B_V-B_R}{B_V+B_R} \tag{31}$$

$$A_G = \frac{G_V - G_R}{G_V + G_R} \tag{32}$$

$$A^U = \frac{B_V}{B_R} + 5\frac{G_V}{G_R} - 6 \geq 0 \tag{33}$$

$$d_E = \sqrt{A^U + 6} \tag{34}$$

$$A^{eq} = \left(1 + \frac{5}{12}A^U\right) + \sqrt{\left(1 + \frac{5}{12}A^U\right)^2 - 1} \tag{35}$$

$$A^L = \sqrt{\left[\ln\left(\frac{B_V}{B_R}\right)\right]^2 + 5\left[\ln\left(\frac{C_{44}^V}{C_{44}^R}\right)\right]^2} \tag{36}$$

with,

$C_{44}^R = \left(\frac{5}{3}\right)\frac{C_{44}(C_{11}-C_{12})}{3(C_{11}-C_{12})+4C_{44}}$ is the Reuss value of $C_{44}$

$C_{44}^V = C_{44}^R + \left(\frac{3}{5}\right)\frac{(C_{11}-C_{12}-2C_{44})^2}{3(C_{11}-C_{12})+4C_{44}}$ is the Voigt value of $C_{44}$

$$A_{B_a} = \frac{B_a}{B_b} = \alpha \tag{37}$$

$$A_{B_c} = \frac{B_c}{B_b} = \frac{\alpha}{\beta} \tag{38}$$

**Table 6.** The Zener anisotropy parameter ($A$), shear anisotropy parameters ($A_1$, $A_2$, and $A_3$), anisotropy in compressibility ($A_B$) and anisotropy in shear ($A_G$) of $X$Ir$_3$ ($X$ = La, Th).

| Compound | $A$ | $A_1$ | $A_2$ | $A_3$ | $A_B$ | $A_G$ |
|---|---|---|---|---|---|---|
| LaIr$_3$ | 0.87 | 0.50 | 0.50 | 1 | 0.0004 | 0.3811 |
| ThIr$_3$ | 0.28 | 0.20 | 0.20 | 1 | 0.0003 | 0.3674 |

The shear anisotropy parameters $A_1$, $A_2$, and $A_3$ correspond to the shear planes {100}, {010}, and {001}, respectively. It is seen from the Table 6 that the values of $A_1$ and $A_2$ are not equal to one which indicates that both the compound are elastically anisotropic in terms of shearing stress on different crystal planes. From the values of $A_1$ and $A_2$, it can be suggested that LaIr$_3$ is elastically less anisotropic than ThIr$_3$. It is also observed that the values of $A_1$ and $A_2$ are equal reflecting the rhombohedral structure of our studied compounds. Moreover, we found that the shear anisotropy factor $A_3$ is equal to unity for both the compounds, suggesting that the compounds are isotropic between the [110] and [010] directions in the {001} shear planes. The values of anisotropy in compressibility ($A_B$) and anisotropy in shear ($A_G$) are equal to zero for an isotropic crystal, while a value of 1 refers to the maximum probable anisotropy of the crystal [57]. For both the materials the calculated value of $A_B$ is much lower than $A_G$, which implies that the anisotropy in compressibility is much lower than the anisotropy in shear.

**Table 7.** The universal anisotropy index ($A^U$, $d_E$), equivalent Zener anisotropy index ($A^{eq}$), universal log-Euclidean index ($A^L$), anisotropies of the bulk modulus along $a$- and $c$-axis ($A_{Ba}$ and $A_{Bc}$) of $X$Ir$_3$ ($X$ = La, Th).

| Compound | $A^U$ | $d_E$ | $A^{eq}$ | $A^L$ | $A_{Ba}$ | $A_{Bc}$ |
|---|---|---|---|---|---|---|
| LaIr$_3$ | 6.16 | 3.49 | 7.00 | 2.42 | 1 | 1.11 |
| ThIr$_3$ | 5.81 | 3.44 | 6.69 | 1.93 | 1 | 0.92 |

The universal anisotropy index ($A^U$) is called 'universal' since it applies to all types of crystal symmetries. For an elastically isotropic crystal, $A^U$ is equal to zero, but any other positive number suggests a level of anisotropy of the crystal [58]. In this study, we found that the value of $A^U$ is much higher than zero (see Table 7) for $X$Ir$_3$ ($X$ = La, Th) compounds, which implies a strong anisotropy of the compounds. Again, from Table 7 it is observed that the value of equivalent Zener anisotropy index ($A^{eq}$) is greater than unity, which also indicates the anisotropic nature of the compounds. For crystalline materials, the universal log-Euclidean index ($A^L$) ranges from 0 to 10.26 ($0 \leq (A^L) \leq 10.26$) while $A^L = 1$ for an elastically isotropic crystal [59]. Both the compounds in this study have the values of $A^L$ greater than one, which indicates a degree of anisotropy of the compounds. Furthermore, the estimated $A_{Ba}$ and $A_{Bc}$ values (see Table 7) indicate that the bulk moduli of our investigated compounds are isotropic along the $a$-axis and anisotropic along the $c$-axis; a consequence of the rhombohedral structure.

For an isotropic crystal, the three-dimensional (3D) directional variation of Young's modulus ($Y$), compressibility ($\beta$), shear modulus ($G$), and Poisson's ratio ($\eta$) should be spherical in shape, and any deviation from this configuration would indicate a level of elastic anisotropy. In this study, we have obtained 2D and 3D visualization of $Y$, $\beta$, $G$, and $\eta$ by using elastic tensor analysis (ELATE) [60]. Figures 2 to 5 illustrate the 2D and 3D plots generated by ELATE. The 3D plots of $Y$, $G$, and $\eta$ show a larger deviation from spherical shape, suggesting a significant amount of elastic anisotropy. On the other hand, the 3D plot of $\beta$ shows a small deviation from spherical configuration, signifying a low level of anisotropy in compressibility. Again, it is also seen from the plots that the elastic parameters are fully isotropic in the $ab$-plane as also indicated by the value of $A_{Ba}$ (Table 7).

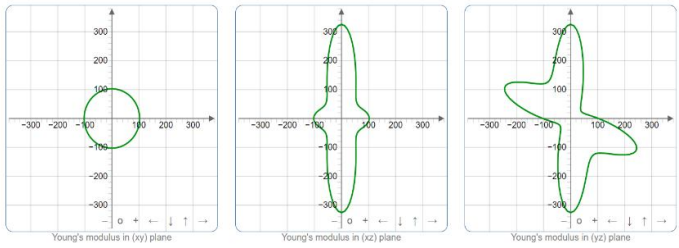 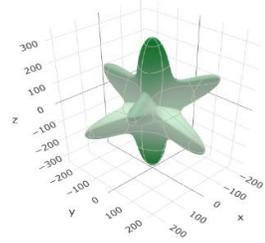

2D visualization of *Y*         3D visualization of *Y*

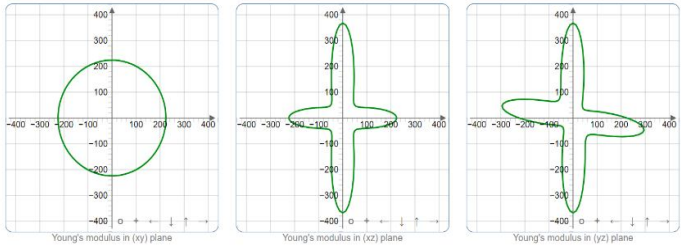 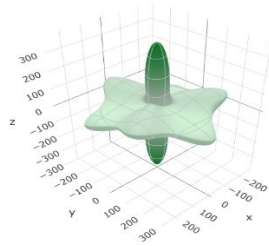

2D visualization of *Y*         3D visualization of *Y*

**Figure 2.** 2D and 3D directional variation of Young's modulus (*Y*) for $X$Ir$_3$ ($X$ = La, Th).

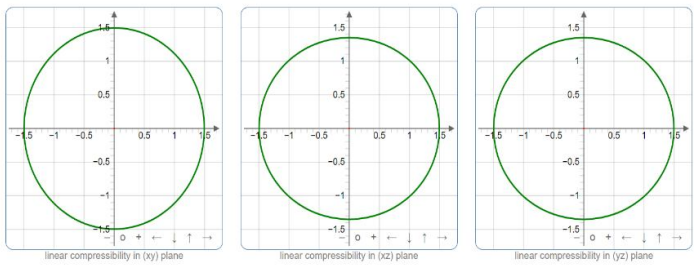 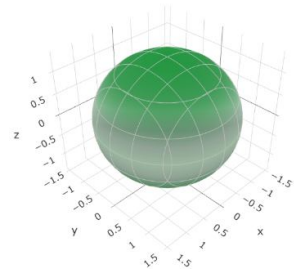

2D visualization of *β*         3D visualization of *β*

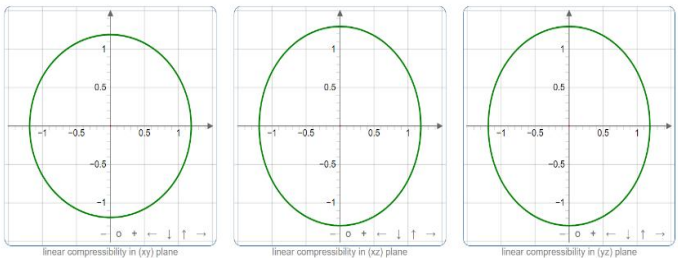 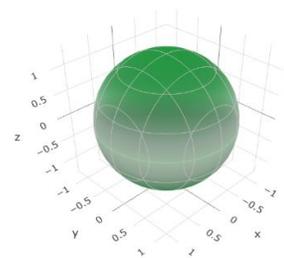

2D visualization of *β*         3D visualization of *β*

**Figure 3.** 2D and 3D directional variation of linear compressibility (*β*) for $X$Ir$_3$ ($X$ = La, Th).

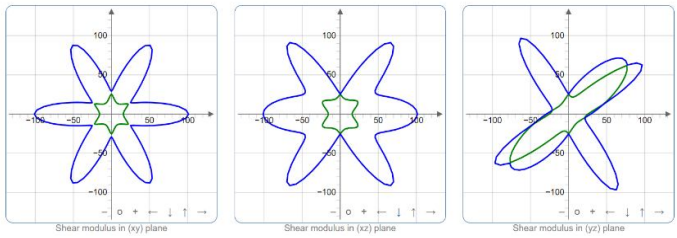 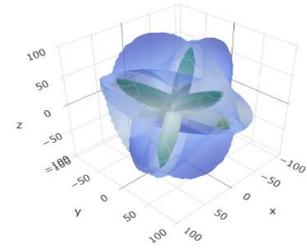

2D visualization of $G$      LaIr$_3$      3D visualization of $G$

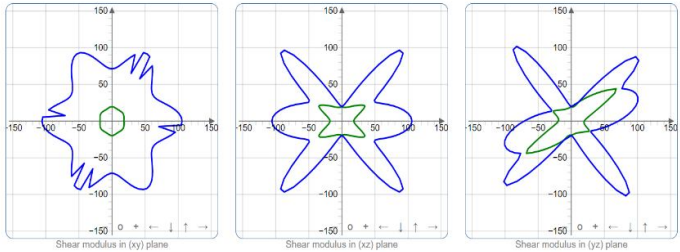 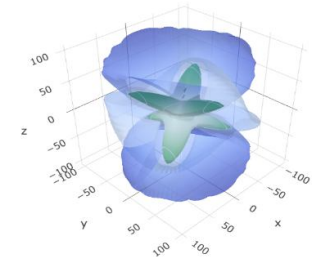

2D visualization of $G$      ThIr$_3$      3D visualization of $G$

**Figure 4.** 2D and 3D directional variation of shear modulus ($G$) for $X$Ir$_3$ ($X$ = La, Th).

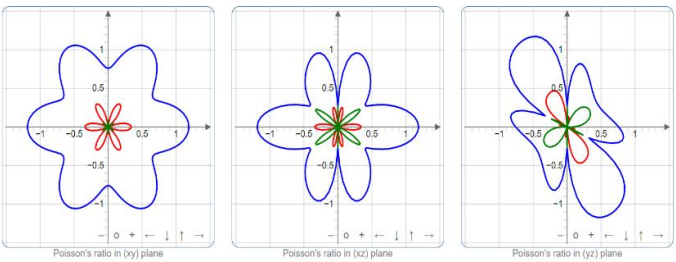 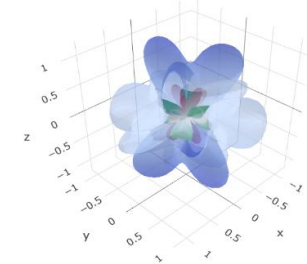

2D visualization of $\eta$      LaIr$_3$      3D visualization of $\eta$

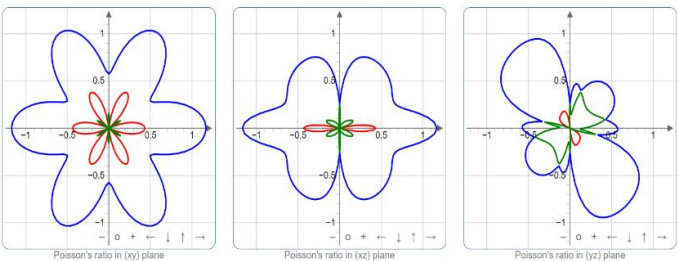 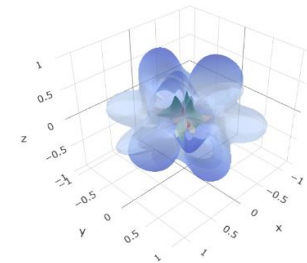

2D visualization of $\eta$      ThIr$_3$      3D visualization of $\eta$

**Figure 5.** 2D and 3D directional variation of Poisson's ratio ($\eta$) for $X$Ir$_3$ ($X$ = La, Th).

*3.3. Mulliken and Hirshfeld population analysis*

The Mulliken bond population analysis is a widely used method in quantum chemistry to comprehend the nature of chemical bonds (ionic, covalent and metallic) in a material. In this section, we have employed both Mulliken population analysis (MPA) [22] and Hirshfeld population analysis (HPA) [61] to explore the bonding natures of our chosen compounds. The results obtained from the bond population analysis (MPA and HPA) of $X$Ir$_3$ ($X$ = La, Th) compounds are disclosed in Table 8.

The charge spilling index determines how much valence charge is missing in a projection. A low value of this parameter suggests that electronic bonding is efficiently represented in the MPA. Table 8 shows that the charge spilling parameters are very low, which indicates that the projections of atomic orbitals are efficiently represented in the MPA.

The analysis also shows that the Mulliken and Hirshfeld charges are positive for the $X$ ($X$ = La, Th) atoms (act as cation) while these charges are negative for the Ir atom (acts as anion). The value of Mulliken charge and Hirshfeld charge (see Table 8) of our studied compounds indicates that electrons are partially transferred from Ir to $X$ atoms. This result implies the presence of ionic bonds the $X$ and Ir atoms. It is also noticed from the Table 8 that the Mulliken charge transfer between La and Ir atoms are higher than the charge transfer between Th and Ir atoms.

The difference between the formal ionic charge and the computed Mulliken charge is known as the effective valence charge (EVC) of an atom [24]. In this study, we have measured the EVC of $X$Ir$_3$ ($X$ = La, Th) compounds in order to predict the degree of covalency or iconicity. EVC values larger than zero imply increasing degree of covalency, while the zero value of EVC denotes an ideal ionic bond. The non-zero values of EVC (see Table 8) implies the existence of both covalent and ionic bonds in $X$Ir$_3$ ($X$ = La, Th) compounds.

**Table 8.** Charge spilling parameter (%), orbital charges (electronic charge), atomic Mulliken charge (electronic charge), EVC (electronic charge), and Hirshfeld charge (electronic charge) of $X$Ir$_3$ ($X$ = La, Th).

| Compound | Charge spilling | Atoms | s | p | d | f | Total | Mulliken charge | EVC | Hirshfeld charge |
|---|---|---|---|---|---|---|---|---|---|---|
| LaIr$_3$ | 0.12 | La | 1.44 | 6.05 | 1.83 | 0.00 | 9.32 | 1.68 | 1.32 | 0.15 |
| | | | 1.66 | 5.96 | 1.99 | 0.00 | 9.61 | 1.39 | 1.61 | 0.11 |
| | | Ir | 0.81 | 0.99 | 7.69 | 0.00 | 9.49 | -0.49 | -2.51 | -0.03 |
| | | | 0.86 | 0.81 | 7.76 | 0.00 | 9.44 | -0.44 | -2.56 | -0.09 |
| | | | 0.84 | 0.97 | 7.70 | 0.00 | 9.51 | -0.51 | -2.49 | -0.04 |
| | | Th | 2.41 | 5.96 | 2.20 | 0.57 | 11.14 | 0.86 | 2.14 | 0.19 |
| | | | 2.49 | 5.80 | 2.37 | 0.64 | 11.30 | 0.70 | 2.30 | 0.11 |

| | | | 0.72 | 0.82 | 7.70 | 0.00 | 9.25 | -0.25 | -2.75 | -0.05 |
|---|---|---|---|---|---|---|---|---|---|---|
| ThIr$_3$ | | Ir | 0.76 | 0.69 | 7.74 | 0.00 | 9.19 | -0.19 | -2.81 | -0.07 |
| | | | 0.75 | 0.82 | 7.69 | 0.00 | 9.26 | -0.26 | -2.74 | -0.04 |

### *3.4. Theoretical bond hardness*

Hardness is a fundamental mechanical property of a material which measures its resistance to deformation, abrasion, or indentation. It also describes the ability of a material to endure localized stresses without considerable plastic deformation or damage. In industry, the relative hardness of a material is also frequently employed. A material's hardness is a crucial physical parameter to comprehend its applicability, particularly when used as an abrasive and radiation tolerant element [62]. In general, a harder material has higher bond population, higher degree of covalency, and shorter bond length. In general, a solid with a higher bulk modulus and shear modulus also has a higher crystal stiffness and hardness. Although the bulk modulus and shear modulus give some information on hardness, there is no straightforward relationship with hardness and bulk or shear modulus [63]. The two forms of hardness are intrinsic and extrinsic. In general, the hardness of a single crystal is regarded as intrinsic, whereas the hardness of polycrystalline materials is considered as extrinsic. The bond hardness is an useful intrinsic measure of a material and can be obtained by using the following expressions [64,65]:

$$H_v^\mu = \left[\prod^\mu \left\{740(P^\mu - P^{\mu'})(v_b^\mu)^{-5/3}\right\}^{n^\mu}\right]^{1/\sum n^\mu} \qquad (39)$$

$$H_v = \left[\prod^\mu (H_v^\mu)^{n^\mu}\right]^{1/\sum n^\mu} \qquad (40)$$

$$P^{\mu'} = \frac{n_{free}}{V_0} \qquad (41)$$

$$n_{free} = \int_{E_P}^{E_F} N(E)\, dE \qquad (42)$$

$$v_b^\mu = \frac{(d^\mu)^3}{\sum_v [(d^\mu)^3 n_b^\mu]} \qquad (43)$$

where, $H_v^\mu$ signifies the bond hardness of the $\mu$-type bond, $H_v$ is the total hardness of the compound, $P^\mu$ and $P^{\mu'}$ are the Mulliken bond overlap population and metallic population, $n_{free}$ is the number of free electrons, $V_0$ is the cell volume, $E_F$ and $E_P$ refer to the energy at the Fermi level and at the pseudo-gap, $n^\mu$ denotes the total number of $\mu$-type bond, $v_b^\mu$ refers to the bond volume of $\mu$-type bond, $d^\mu$ is the bond length of $\mu$-type bond, and $n_b^\mu$ refers to the $\mu$-type bond density per cubic angstroms.

The metallic nature of bonds in a solid can also be estimated from the Mulliken bond population analysis. In this study, we have calculated the metallicity ($f_m$) of $X$Ir$_3$ ($X$ = La, Th) compounds by using the following equations [66]:

$$f_m = \frac{P^{\mu'}}{P^\mu} \tag{44}$$

The calculated number of free electrons, bond length, Mulliken and metallic population, total number of each type of bonds, metallicity, bond volume and bond hardness of each type of bonds and the total hardness of our investigated compounds are listed in Table 9. The Mulliken bond populations of a compound indicate the degree of overlapping of the electron clouds generating bonds between the compound's two relevant atoms. For a purely ionic bond, the value of overlap population is equal to zero but any other positive value indicates a level of covalency of the bond. The bonding and anti-bonding kind of interactions between the relevant pairs of atoms can be identified from the positive and negative values of a compound's Mulliken bond overlap populations, respectively [67]. From Table 9 it is seen that, both the studied compounds have positive values of bond populations for the Ir-Ir bonding which implies the bonding type of interaction between these atoms. Metallic populations for $X$Ir$_3$ ($X$ = La, Th) compounds are found to be very low (see Table 9). This result suggesting the existence of weak metallic bond in the compounds. It is also noticed from the Table 9 that LaIr$_3$ compound has a higher total bond hardness value than ThIr$_3$ compound. This result is contradictory with the results obtained in the elastic and mechanical properties section. **Discuss in conclusion**

**Table 9.** The calculated number of free electrons ($n_{free}$), bond length ($d^\mu$ in Å), bond number ($n^\mu$), Mulliken bond overlap population ($P^\mu$), metallic population ($P^{\mu'}$), metallicity ($f_m$), bond volume ($v_b^\mu$ in Å$^3$), bond hardness of the $\mu$-type bond ($H_v^\mu$ in GPa), and total hardness ($H_v$ in GPa) of $X$Ir$_3$ ($X$ = La, Th) compounds.

| Compound | $n_{free}$ | Bond | $d^\mu$ | $n^\mu$ | $P^\mu$ | $P^{\mu'}$ | $f_m$ | $v_b^\mu$ | $H_v^\mu$ | $H_v$ |
|---|---|---|---|---|---|---|---|---|---|---|
| LaIr$_3$ | 40.08 | Ir-Ir | 2.61 | 18 | 0.43 | 0.06 | 0.14 | 8.57 | 7.63 | 7.36 |
| | | | 2.64 | 18 | 0.35 | | 0.17 | 8.91 | 5.60 | |
| | | | 2.65 | 18 | 0.80 | | 0.08 | 8.97 | 14.14 | |
| | | | 2.74 | 18 | 0.36 | | 0.17 | 9.91 | 4.86 | |
| ThIr$_3$ | 21.05 | Ir-Ir | 2.64 | 18 | 0.55 | 0.03 | 0.05 | 8.69 | 10.48 | 5.05 |
| | | | 2.65 | 18 | 0.25 | | 0.12 | 8.74 | 4.39 | |
| | | | 2.65 | 18 | 0.23 | | 0.13 | 8.81 | 3.94 | |
| | | | 2.70 | 18 | 0.23 | | 0.13 | 9.29 | 3.60 | |

*3.5. Electronic charge density distribution*

The electronic charge density distribution map illustrates detailed information about the bonding nature and amount of charge transfer among the atoms of a compound. To estimate the bonding nature of the $X$Ir$_3$ ($X$ = La, Th) compounds, we have computed the electronic charge density distribution maps within the (110) and (111) crystal planes using LDA functionals and the obtained results are depicted in Figure 6. The total electron density is shown by the color scale placed between the panels. In the map, high charge density is indicated by the blue color and low charge density is indicated by the red color. From Figure 6 it is seen that, the electron density around the $X$ ($X$ = La, Th) atoms is much higher compared to the Ir atoms. Since the charge density is found to be almost spherical around the $X$ atoms (see Figure 6), it can be concluded that no charge sharing occurs between the $X$ and the Ir atoms. Thus, the bonding among the $X$-Ir atoms is of ionic nature. Again, charge overlapping is seen between the relevant Ir atoms of $X$Ir$_3$ compounds. This point towards the covalent bonding nature in the crystal. Therefore, both ionic and covalent bonds coexist in $X$Ir$_3$ compounds where the ionic bond dominates. Far from the atomic sites, we have found uniform low density charge distribution in these compounds. This result suggests the presence of a weak metallic bond in these compounds. The overall outcomes in this section are well matched with the bond population analysis that was previously addressed.

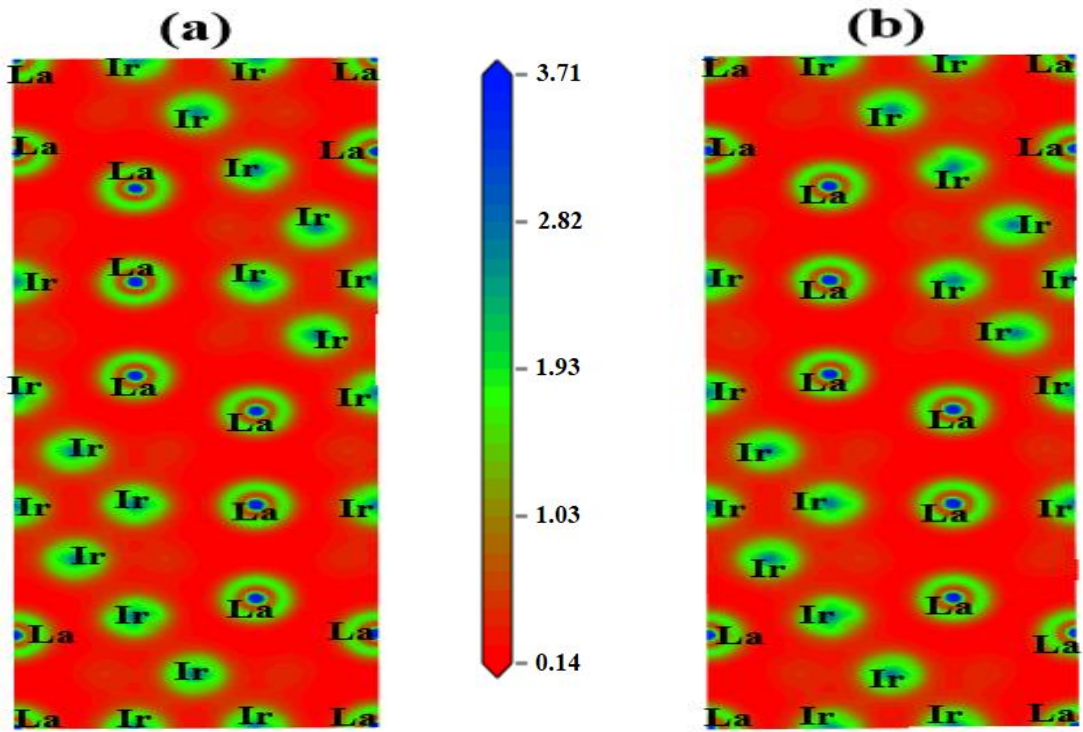

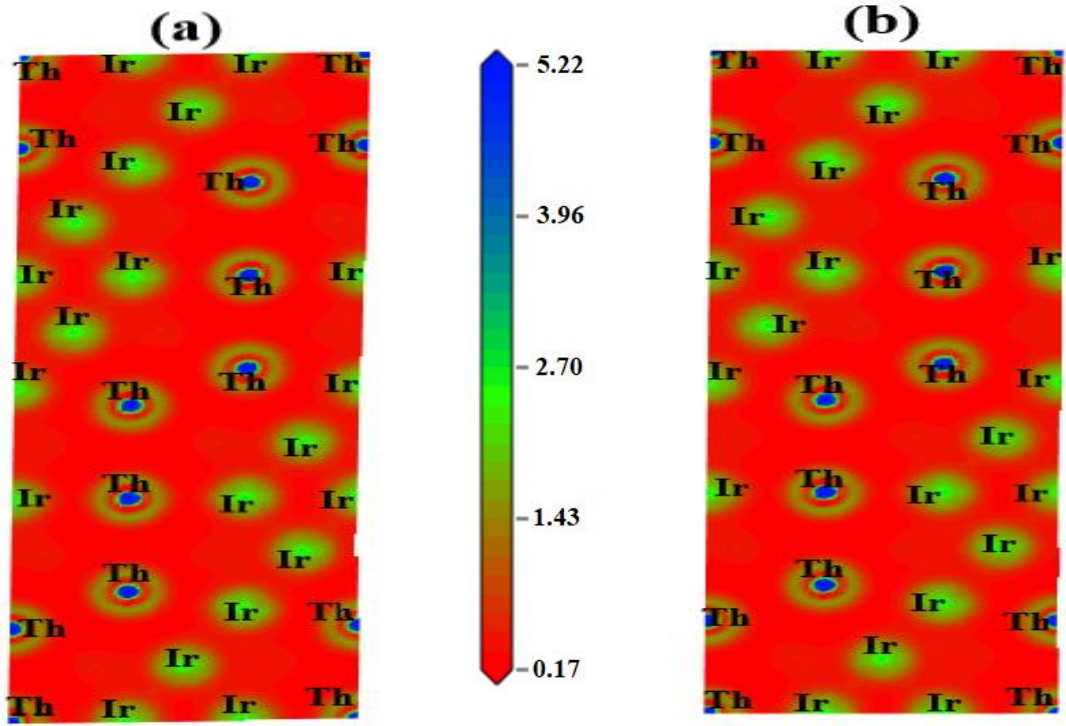

**Figure 6.** The electronic charge density map in the (a) (110) and (b) (111) planes of $X$Ir$_3$ ($X$ = La, Th) compounds. The scale for the charge density map is given between the panels.

*3.6. Electronic properties*

*(a) Electronic band structure*

One of the most significant topics of condensed matter physics is the study of electronic band structure, which is strongly linked to a compound's electrical conductivity, thermopower, heat capacity, bonding features, Hall effect, optoelectronic properties, magnetic properties, and superconductivity. The electronic energy band structures of optimized $X$Ir$_3$ ($X$ = La, Th) compounds have been calculated along the path Γ-A-H-K-Γ-M-L-H in the first Brillouin zone (BZ). The results obtained from the band structure calculations are depicted in Figure 7. It is seen from the Figure 7 that the band structures of $X$Ir$_3$ ($X$ = La, Th) compounds are quite similar to each other. The horizontal broken line at zero energy represents the Fermi level ($E_F$). LaIr$_3$ and ThIr$_3$ compounds have a total number of 469 and 468 electronic energy bands, respectively. It is evident from the Figure 7 that there is no band gap at the Fermi level since considerable overlapping occurs between the valence and conduction bands. The result suggests that both the compounds have metallic character. The red lines in Figure 7 indicate the bands that cross the Fermi level. There are six LaIr$_3$ bands (bands 170, 171, 172, 173, 174 and 175) and five ThIr$_3$ bands (bands 174, 175, 176, 177 and 178) which cross the Fermi level. In the BZ, these bands have both electron-like and hole-like nature in different directions. It is noticed from the Figure 7 that for both the compounds, the bands near the Γ-point are highly dispersive which suggests high mobility and low effective mass of the charge carriers [68–70]. It is also observed from Figure 7 that the bands around the Fermi level of ThIr$_3$ compound are flatter than the LaIr$_3$ compound. Flat bands give rise to high electron effective mass and stronger electronic correlations.

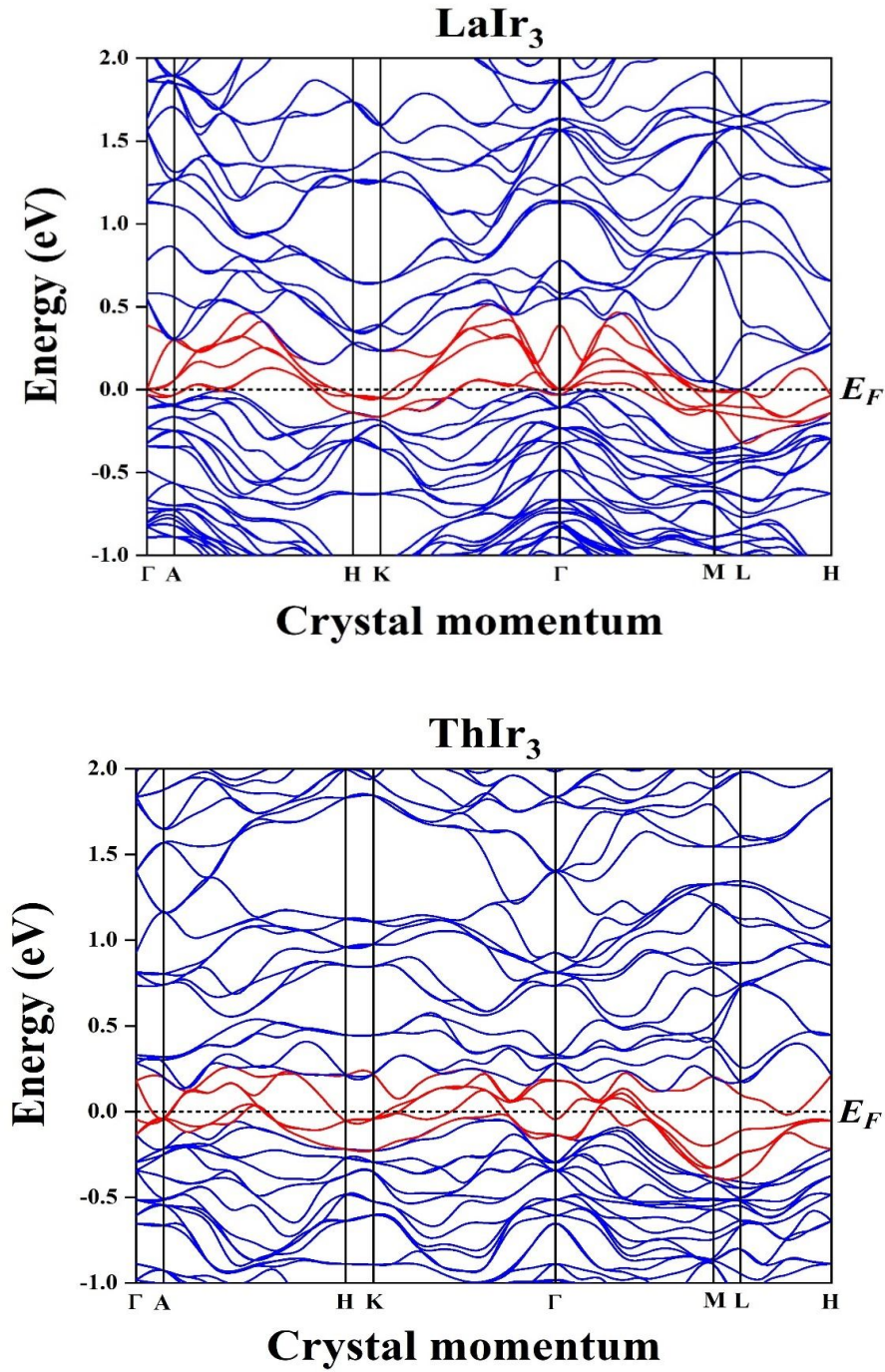

**Figure 7.** The electronic band structure of LaIr$_3$ and ThIr$_3$ compounds along several high symmetry directions in the first BZ.

## (b) Electronic energy density of states (DOS)

The electronic energy density of states (DOS) of a crystalline material is linked with the electronic, optical, superconducting, and magnetic features of a material. The DOS of a material is also crucial to understand how each atom contributes to bonding and anti-bonding states. In this study, we have computed the total and partial density of states (TDOS and PDOS, respectively) of $X$Ir$_3$ ($X$ = La, Th) to gain a better understanding of the electronic features. The results obtained from the TDOS and PDOS calculations of our chosen materials are depicted in Figure 8. The vertical broken line at zero energy represents the Fermi level ($E_F$). From figure 8 it is seen that the TDOS values at the Fermi level are finite. This result confirms again the metallic nature of the compounds.

At $E_F$, the TDOS values of the investigated compounds are quite large (see Figure 8). This suggests that the compounds should exhibit high electrical conductivity. The TDOS values at $E_F$ of LaIr$_3$ and ThIr$_3$ compounds are found to be 31.12 states/eV-unit cell (3.46 states/eV-formula unit) and 28.88 states/eV-unit cell (3.21 states/eV-formula unit), respectively. The value of the TDOS of a compound at the Fermi level, $N(E_F)$, is an important parameter for the purpose of determining electronic stability of the compound. Compounds with lower $N(E_F)$ values are more stable compared to those with higher $N(E_F)$ values [71,72]. As a result, it can be estimated that ThIr$_3$ will be more electronically stable than LaIr$_3$. From previous study, the values of TDOS were found to be 3.90 states/eV-formula unit [5] for LaIr$_3$ and 3.46 states/eV-formula unit [4] for ThIr$_3$. Thus, the results obtained in this study are in good agreement with the previous values.

In order to understand the contribution of $X$ ($X$ = La, Th) and Ir atoms in the TDOS separately, we have computed the PDOS of each atom at different energies. It is seen from Figure 8 that the Ir-$5d$ orbitals of the two compounds under study largely contribute to the TDOS in the proximity of $E_F$. Thus, the metallic conductivity in $X$Ir$_3$ ($X$ = La, Th) compounds are governed by the $5d$ states of Ir atom. Moreover, the compound LaIr$_3$ will exhibit more metallic nature than ThIr$_3$ since the Ir-$5d$ states at $E_F$ is stronger in LaIr$_3$ than in ThIr$_3$. The valence band of LaIr$_3$ compound is mainly composed by the $5d$ orbital of Ir with a modest input from the La-$5d$ orbital. On the other hand, the conduction band of LaIr$_3$ is mainly formed by the La-$5d$ states with some contribution from the Ir-$5d$ states. Again, for the ThIr$_3$ compound, the Ir-$5d$ orbital dominates the valence band, while above $E_F$, the Th-$5f$ orbital mainly contribute to the large DOS peak. From Figure 8, it can be inferred that the La-$6s$, Th-$7s$ and Ir-$6s$ states have no effect on the DOS at the $E_F$.

The existence of a deep valley in the TDOS plot close to the $E_F$ is referred to a pseudogap or quasi-gap. It is linked to a material's electronic stability and distinguishes the bonding states from the antibonding states [71–73]. The bonding peak is defined as the nearest negative energy peak of TDOS in the vicinity of $E_F$ and the anti-bonding peak is defined as the nearest positive energy peak of TDOS in the vicinity of $E_F$. It can be inferred from the Figure 8 that for $X$Ir$_3$ ($X$ = La, Th) compounds the bonding and anti-bonding peaks are positioned within 2 eV apart from the $E_F$.

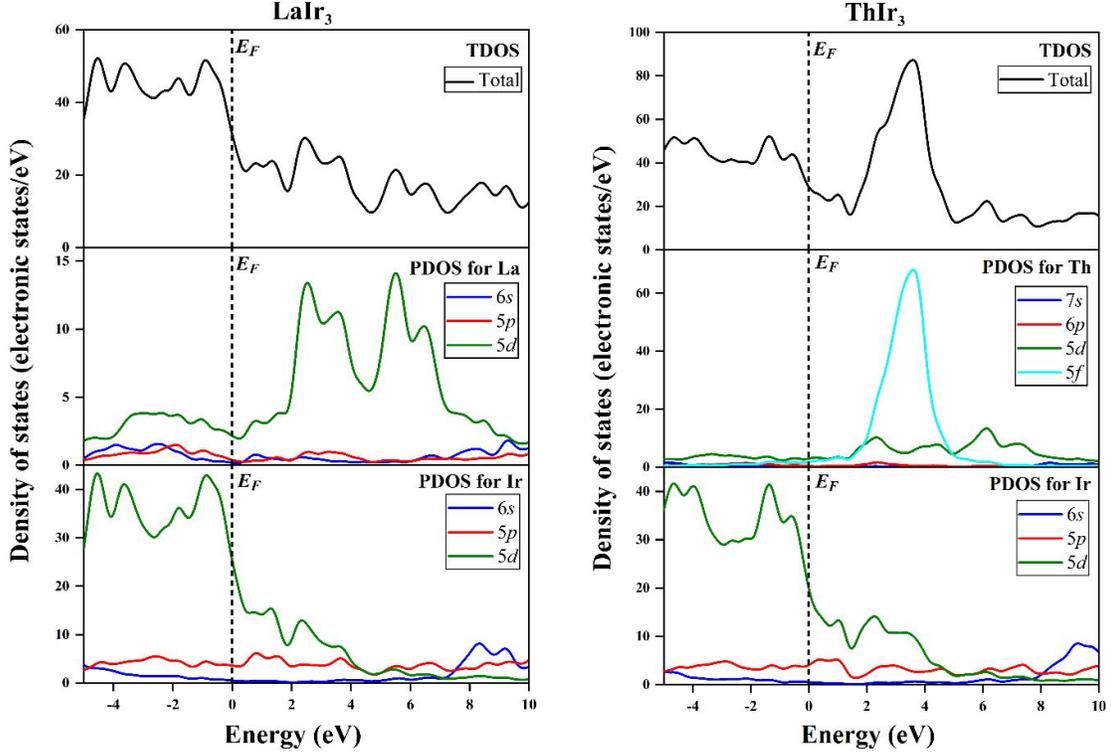

**Figure 8.** The total and partial density of states (TDOS and PDOS, respectively) plots of $X$Ir$_3$ ($X$ = La, Th) as a function of energy.

### 3.7. Thermal properties

*(a) Debye temperature and acoustic velocities*

In solid-state physics, the Debye temperature ($\theta_D$) of a material determines the highest energy mode of lattice vibration and acts as a cutoff for the phonon energy distribution. This parameter also gives useful information on the lattice dynamics, specific heat capacity, thermal conductivity, melting temperature, and elastic constants of a material. In this study, we have employed the Anderson method [74] to calculate the $\theta_D$. This method uses the sound velocity ($v_a$) to calculate $\theta_D$ via the following equation:

$$\theta_D = \frac{h}{k_B}\left[\left(\frac{3n}{4\pi V_0}\right)\right]^{\frac{1}{3}} v_a \tag{45}$$

In the above equation, $V_0$ signifies the cell volume, $n$ defines the total number of atoms in the cell, $h$ refers to the Planck's constant and $k_B$ denotes the Boltzmann's constant. The following expression can be used to calculate the average sound velocity ($v_a$) [75]:

$$v_a = \left[\frac{1}{3}\left(\frac{2}{v_t^3} + \frac{1}{v_l^3}\right)\right]^{-\frac{1}{3}} \tag{46}$$

here, $v_l$ and $v_t$ refer to the longitudinal and transverse sound velocities, respectively. The $v_l$ and $v_t$ can be obtained from:

$$v_l = \sqrt{\frac{3B + 4G}{3\rho}} \tag{47}$$

$$v_t = \sqrt{\frac{G}{\rho}} \tag{48}$$

where, $\rho$ refers to the density of the solid. The calculated values of $\theta_D$ of $X$Ir$_3$ ($X$ = La, Th) compounds along with the mass density ($\rho$) and acoustic velocities ($v_l$, $v_t$ and $v_a$) are presented in Table 10.

It is seen from the table 10 that the calculated $v_a$ and $\theta_D$ of LaIr$_3$ are considerably larger than that of ThIr$_3$. This happens due to the lower mass density of LaIr$_3$ compounds than the ThIr$_3$ compounds. Moreover, the values of $v_t$ are found to be much lower than $v_l$ (see Table 10) in $X$Ir$_3$ ($X$ = La, Th) compounds. The overall $\theta_D$ of $X$Ir$_3$ compounds are found to be low. This result suggests a low thermal conductivity of the compounds which is further discussed in the subsequent section. It is also worth noting that the calculated value of $\theta_D$ for LaIr$_3$ is significantly lower than the previously anticipated experimental result [1]. However, the calculated value of $\theta_D$ for ThIr$_3$ is found to be significantly higher than the experimental result [4].

**Table 10.** The calculated density ($\rho$ in kg/m$^3$), acoustic velocities ($v_l$, $v_t$ and $v_a$, all in m/sec) and Debye temperature ($\theta_D$ in K) of $X$Ir$_3$ ($X$ = La, Th) along with the previously estimated experimental (Expt.) values.

| Compound | $\rho$ | $v_l$ | $v_t$ | $v_a$ | $\theta_D$ | Ref. |
|---|---|---|---|---|---|---|
| LaIr$_3$ | 16345.59 | 4283.65 | 1785.33 | 2019.61 | 229 | This work |
|  | - | - | - | - | 366 | [1]$^{Expt.}$ |
| ThIr$_3$ | 18920.58 | 4297.49 | 1747.22 | 1978.15 | 226 | This work |
|  | - | - | - | - | 169 | [4]$^{Expt.}$ |

*(b) Anisotropies in acoustic velocity*

A solid's acoustic velocity is an essential feature that is linked to its thermal and electrical properties. A solid's atoms produce three vibrational modes: two transverse ($v_{t1}$ and $v_{t2}$) modes and one longitudinal ($v_l$) mode. These vibrational modes are excited by sound wave when it travels through a solid. The presence of elastic anisotropy in a solid can be detected by the anisotropy in sound velocities, and vice versa. For rhombohedral crystal structure, the pure longitudinal and transverse modes can be determined only along [100] and [001] crystallographic directions. In this work, we have measured the acoustic velocities of $X$Ir$_3$ ($X$ = La, Th) compounds along the [100] and [001] propagation directions in order to comprehend the anisotropic nature of acoustic velocities. We have employed the following equations [76] to calculate the acoustic velocities and the outcomes are presented in Table 11.

**[100] direction:**

$$[100]v_l = \sqrt{(C_{11} - C_{12})/2\rho} \ ; \ [010]v_{t1} = \sqrt{C_{11}/\rho} \ ; \ [001]v_{t2} = \sqrt{C_{44}/\rho} \qquad (49)$$

**[001] direction:**

$$[001]v_l = \sqrt{C_{33}/\rho}; \ [100]v_{t1} = [010]v_{t2} = \sqrt{C_{44}/\rho} \qquad (50)$$

here, $v_l$ denotes the longitudinal mode, $v_{t1}$ refers to the first transverse mode and $v_{t2}$ denotes the second transverse mode.

Table 11 shows that the sound velocities of $X$Ir$_3$ materials along [100] and [001] directions are not same which implies that the materials will have lattice dynamical anisotropy. Thus, we can estimate that our materials will exhibit direction-dependent thermal and charge transport properties.

**Table 11.** Anisotropic acoustic velocities (m/s) in $X$Ir$_3$ ($X$ = La, Th) along various propagation directions.

| Propagation directions | | LaIr$_3$ | ThIr$_3$ |
|---|---|---|---|
| [100] | $[100]v_l$ | 1955.65 | 2265.04 |
|  | $[010]v_{t1}$ | 4488.65 | 4735.82 |
|  | $[001]v_{t2}$ | 1825.82 | 1189.04 |
| [001] | $[001]v_l$ | 5005.08 | 4917.16 |
|  | $[100]v_{t1}$ | 1825.82 | 1189.04 |
|  | $[010]v_{t2}$ | 1825.82 | 1189.04 |

Except the $[100]v_l$ mode, all the other sound velocities are higher in LaIr$_3$ compared to those in ThIr$_3$.

*(c) Lattice thermal conductivity*

The lattice thermal conductivity ($k_{ph}$) can be defined as the measurement of the ability of a material to conduct heat via phonons through its crystal structure. It demonstrates how heat is efficiently conducted through a material as a result of the phonons propagation. High thermal conductivity materials are desirable for efficient heat dissipation for applications such as in electronic devices, whereas low thermal conductivity materials are useful for thermal barriers. In this study, we have calculated the $k_{ph}$ of $X$Ir$_3$ ($X$ = La, Th) compounds at room temperature (300K) by employing the following empirical formula proposed by Slack [77] and the obtained values are disclosed in Table 12:

$$k_{ph} = A(\gamma) \frac{M_{av} \theta_D^3 \delta}{\gamma^2 n^{2/3} T} \tag{51}$$

In the above formula, $\theta_D$ is the Debye temperature, $M_{av}$ stands for the average atomic mass, $\delta$ refers to the cubic root of the average atomic volume, $n$ denotes the total number of atoms in the unit cell, $\gamma$ is the Grüneisen parameter and $T$ stands for the absolute temperature. The following relations [77,78] can be employed to obtain the parameter $\gamma$ and the factor $A(\gamma)$:

$$\gamma = \frac{3(1 + \eta)}{2(2 - 3\eta)} \tag{52}$$

$$A(\gamma) = \frac{5.720 \times 10^7 \times 0.849}{2 \times (1 - 0.514/\gamma + 0.228/\gamma^2)} \tag{53}$$

where, $\eta$ stands for the Poisson's ratio.

It is observed from the Table 12 that the lattice thermal conductivities of $X$Ir$_3$ ($X$ = La, Th) materials are extremely low. The low values of $k_{ph}$ and $\theta_D$ of $X$Ir$_3$ materials suggesting that they can be used for thermal insulation purpose. It is also found that the $k_{ph}$ value of LaIr$_3$ is larger than that of ThIr$_3$. Therefore, the compound LaIr$_3$ is more thermally conductive than the ThIr$_3$ compound.

*(d) Minimum thermal conductivity and its anisotropy*

The lowest possible phonon thermal conductivity that a material exhibits at high temperatures is defined as the minimum thermal conductivity ($k_{min}$) of the material. The investigation of a material's minimal thermal conductivity has played vital role in the identification and development of material that can be utilized as TBC (thermal barrier coating) material. The presence of defects in a crystal has no effect on the $k_{min}$. In this work, we have calculated the $k_{min}$ of $X$Ir$_3$ materials by employing the following formula proposed by Clarke [79]:

$$k_{min} = k_B v_a (V_{atomic})^{-\frac{2}{3}} \tag{54}$$

here, $V_{atomic}$ refers to the atomic volume, $v_a$ denotes the average sound velocity and $k_B$ is the Boltzmann constant.

The transfer of heat via thermal vibrations is dominated by the propagation of elastic waves. Since both the compounds under study are elastically anisotropic, they will have anisotropy in minimum thermal conductivity. The $k_{min}$ of $X$Ir$_3$ materials along different crystallographic directions can be obtained by using the estimated acoustic velocities of the materials. In this study, we have employed the Cahill and Clarke [80] expression to estimate the $k_{min}$ of $X$Ir$_3$ materials along [100] and [001] directions.

$$k_{min} = \frac{k_B}{2.48} n^{2/3} (v_l + v_{t1} + v_{t2}) \tag{55}$$

where, $n$ denotes the number of atoms per mole. The values of $k_{min}$ estimated from the Cahill and Clarke model are presented in Table 12.

**Table 12.** Lattice thermal conductivity ($k_{ph}$ in W/m-K) evaluated by Slack method at 300 K, number of atoms per unit volume ($n$ in m$^{-3}$), and minimum thermal conductivity ($k_{min}$ in W/m-K) evaluated by Cahill and Clarke model along [100] and [001] crystallographic directions of $X$Ir$_3$ ($X$ = La, Th) compounds.

| Compound | $k_{ph}$ | $n$ (10$^{28}$) | [100] $k_{min}$ | [001] $k_{min}$ | $k_{min}$ | |
|---|---|---|---|---|---|---|
| | | | | | Cahill | Clarke |
| LaIr$_3$ | 0.80 | 5.50 | 0.67 | 0.70 | 0.63 | 0.40 |
| ThIr$_3$ | 0.78 | 5.63 | 0.67 | 0.60 | 0.64 | 0.39 |

Table 12 shows that the values of $k_{min}$ for both the materials are equal in [100] direction while in [001] direction the $k_{min}$ of LaIr$_3$ is larger than that of ThIr$_3$. Different values of $k_{min}$ in deferent directions indicate the presence of thermal anisotropy in the materials. It is also noted that the Cahill model estimates slightly higher $k_{min}$ than the Clarke model (see Table 12). It is noted that a material should have a $k_{min}$ value equal to or less than 1.25 W/m-K to be used as suitable TBC material [43,81]. Therefore, the investigated materials can be used as potential TBC material.

*(e) Thermal expansion coefficient, heat capacity and wavelength of the dominant phonon*

A material's thermal expansion coefficient ($\alpha$) is a physical parameter that measures how much its dimensions change in response to a change in temperature. The potential of a material to be employed as a TBC material is strongly correlated with this parameter. This parameter is also linked to a variety of other physical properties, such as the thermal conductivity, melting temperature, specific heat, and so on. It is known that the α of a material is inversely related to its melting temperature ($T_m$), as follows: $\alpha \approx 0.02 / T_m$ [79,82].

Another essential thermophysical parameter is the heat capacity per unit volume ($\rho C_P$). This parameter is also known as volumetric heat capacity. The change in heat energy of a unit volume in a material by one degree Kelvin change in temperature is called the $\rho C_P$ of the material. It measures a material's capacity to store thermal energy as well as its thermal response to temperature changes.

Phonons are responsible for the heat and sound energy transmission in solids. It is defined as the quanta of lattice vibrations. Phonon plays a crucial role to determine the thermal conductivity, heat capacity and many other thermal and electrical properties. The wavelength corresponding to the peak of the phonon distribution function is known as the dominant phonon wavelength ($\lambda_{dom}$).

In this section, we have estimated the $\alpha$, $\rho C_P$ and $\lambda_{dom}$ of $X$Ir$_3$ ($X$ = La, Th) by using the relationships [79,82,83] given below and the results are disclosed in Table 13.

$$\alpha = \frac{1.6 \times 10^{-3}}{G} \tag{56}$$

$$\rho C_P = \frac{3k_B}{\Omega} \tag{57}$$

$$\lambda_{dom} = \frac{12.566\, v_a}{T} \times 10^{-12} \tag{58}$$

In the above expressions, $G$ is the shear modulus, $\Omega$ denotes the volume per atom, and $v_a$ is the average sound velocity. It is seen from the Table 13 that the values of $\alpha$ and $\lambda_{dom}$ are much higher in LaIr$_3$ than that of ThIr$_3$. This happens due to the fact that the compound LaIr$_3$ has smaller shear modulus and larger sound velocity compared to those of the ThIr$_3$ compound. On the other hand, the value of $\rho C_P$ is found to be higher in ThIr$_3$ compared to the LaIr$_3$.

*(f) Melting temperature*

A material's melting temperature, also known as its melting point, is the temperature at which it transforms from a solid to a liquid phase at a certain atmospheric pressure. Nowadays the melting temperature ($T_m$) of a material is a fascinating and necessary topic of research. Higher melting temperature materials have lower thermal expansion, higher cohesive energy, and higher bonding energy [83]. Using the optimized elastic constants, we have calculated the $T_m$ of $X$Ir$_3$ compounds by employing the formula given below [84]:

$$T_m = 354 + 1.5(2C_{11} + C_{33}) \tag{59}$$

The calculated values of $T_m$ of $X$Ir$_3$ materials are listed in Table 13. It is found that both the studied compounds have high melting point. This result suggests that they are suitable for high-temperature applications. It is also noticed that the compound ThIr$_3$ has much higher melting point compared to the LaIr$_3$ compound. Thus, ThIr$_3$ will have higher bonding and cohesive energy compared to LaIr$_3$.

**Table 13.** The calculated thermal expansion coefficient ($\alpha$ in K$^{-1}$), volumetric heat capacity ($\rho C_P$ in JK$^{-1}$m$^{-3}$), dominant phonon wavelength at 300K ($\lambda_{dom}$ in m) and melting point ($T_m$ in K) of $X$Ir$_3$ ($X$ = La, Th) materials.

| Compound | α (10⁻⁵) | $\rho C_P$ (10⁶) | $\lambda_{dom}$ (10⁻¹¹) | $T_m$ |
|---|---|---|---|---|
| LaIr$_3$ | 3.07 | 2.28 | 8.46 | 1956.20 |
| ThIr$_3$ | 2.77 | 2.33 | 8.28 | 2313.26 |

### *3.8. Optical properties*

The optical properties of a material are essential for understanding how the material interacts with light and how light is reflected, refracted, absorbed, scattered, and transmitted as it travels through it. A number of energy-dependent optical parameters, including the absorption coefficient $\alpha(\omega)$, reflectivity $R(\omega)$, optical conductivity $\sigma(\omega)$, refractive index $n(\omega)$, dielectric function $\varepsilon(\omega)$, and energy loss function $L(\omega)$, can be used to predict the overall optical response of a material. In this study, we have calculated these optical parameters for $X$Ir$_3$ ($X$ = La, Th) compounds and the outcomes are depicted in Figures 9 and 10. These optical constants have significant consequences for a variety of practical applications, such as optoelectronics, optical data storage, display systems, solar energy conversion, materials research etc. From the previously calculated elastic properties, it was found that the two compounds under study are elastically anisotropic. This suggests that the optical constants of $X$Ir$_3$ compounds might also show anisotropic nature. In this study, we have computed $\alpha(\omega)$, $R(\omega)$, $\sigma(\omega)$, $n(\omega)$, $\varepsilon(\omega)$, and $L(\omega)$ of $X$Ir$_3$ along [100] and [001] polarization directions for photon energies up to 40 eV. Since both the compounds under study are metallic in nature, Drude damping must be included while analyzing the optical parameters [85,86]. Therefore, we have taken a Drude damping of 0.05 eV for the analysis of the optical parameters. The obtained results are discussed in the following sub-sections.

### *(a) Absorption coefficient*

The absorption coefficient ($\alpha$) of a material determines the amount of photon energy absorbed by the material. From the absorption spectra, we can estimate the electronic nature of a material (metal, semiconductor or insulator). This parameter also provides information on the optimal solar energy conversion efficiency. The absorption spectra of $X$Ir$_3$ compounds are shown in Figures 9(a), and 10(a). These plots show that absorption begins from zero photon energy for both the compounds. This confirms the metallic nature of the compounds. It is also noticed that the $\alpha(\omega)$ are quite high in the ultraviolet region (UV) from 20 eV to 30 eV. In this energy range the peak value of $\alpha(\omega)$ of $X$Ir$_3$ is slightly higher in [001] direction compared to that in [100] direction which indicates small optical anisotropy of the compounds. It is also important to note that the α(ω) of $X$Ir$_3$ compounds reduces drastically around 35 eV, which is closely associated with the location of the loss peak.

### *(b) Reflectivity*

The computed frequency dependent optical reflectivity ($R$) profiles of $X$Ir$_3$ compounds along

[100] and [001] directions are depicted in Figures 9(b) and 10(b). For both the compounds, the calculated $R(\omega)$ along [100] and [001] directions have roughly the similar shape, but the peak heights and positions differ slightly. It is also found that the maximum value (about 70% to 78%) of $R(\omega)$ occurs at zero photon energy. The value of $R(\omega)$ then decreases drastically in the infrared and visible energy region (see Figures 9(b) and 10(b)). In the mid UV region, $R(\omega)$ increases abruptly and reaches a peak then decreases sharply around 35 eV.

*(c) Optical conductivity*

The conductivity ($\sigma$) spectra of a material can be used to understand its electromagnetic response in the presence of a time-varying electric field. This parameter also provides important details regarding the electronic band structure and photoconductivity of a material. Figures 9(c) and 10(c) show that the $\sigma(\omega)$ of $X$Ir$_3$ compounds begins from 0 eV which indicates once again the metallic nature of the compounds. The maximum values of $\sigma(\omega)$ of both the studied compounds occur in the lower photon energy (about 1 to 5 eV). Again, the values of $\sigma(\omega)$ tends to zero in the higher photon energy (about 35 eV). Slight change is observed in $\sigma(\omega)$ spectra with respect to the [100] and [001] polarization directions which implies small amount of anisotropy in $X$Ir$_3$. This anisotropy is slightly higher in ThIr$_3$ compound compared to that in the LaIr$_3$ compound.

*(d) Refractive index*

The real (Re), $n(\omega)$ and imaginary (Im), $k(\omega)$ parts of refractive index are two important optical constants. The imaginary part $k(\omega)$ is also known as the extinction coefficient. The $n(\omega)$ of refractive index determines the phase velocity of an electromagnetic wave and the amount of bending of light, while the extinction coefficient governs the amount of absorption or attenuation of light within the medium. For the optimal design of photoelectric devices, accurate knowledge of the complex refractive index of a material is necessary. The variation of $n(\omega)$ and $k(\omega)$ of $X$Ir$_3$ materials for [100] and [001] polarizations of the incident electric field are shown in Figures 9(d) and 10(d). The $n(0)$ (static) values of $X$Ir$_3$ materials are quite high (between 10 and 14), indicating that these materials have the potential to be utilized in applications involving display devices and wave-guides. It can be concluded from the plots that for both the materials, $n(\omega)$ decreases sharply in the lower energy region (0 to 3 eV) and then almost have the same value in the higher energy region. Again, the extinction coefficient ($k$) of $X$Ir$_3$ materials reaches its peak value in the lower energy region. Furthermore, the value of $k(\omega)$ decreases gradually in the mid-ultraviolet region and then tends to zero in the higher energy region (about 35 eV).

*(e) Dielectric function*

The complex dielectric function, $\varepsilon(\omega)$ of a material can be used to calculate the other optical constants, including optical conductivity, refractive index, reflectivity, energy loss function and absorption of light. The real (Re), $\varepsilon_1(\omega)$ part of the dielectric function describes the polarization of the material. This part is also connected to the permittivity component. On the other hand, the imaginary (Im), $\varepsilon_2(\omega)$ part of the dielectric function corresponds to the dielectric loss in the optical system. Figures 9(e) and 10(e) illustrate the variation of $\varepsilon_1(\omega)$ and $\varepsilon_2(\omega)$ of $X$Ir$_3$ compounds as a function of photon energy. From these plots it is seen that the $\varepsilon_1(\omega)$ of $X$Ir$_3$ crosses zero from below in the mid UV range (about 14 to 16 eV), suggesting that the compounds are metal. It is also noticed that for both the compounds the peak value of $\varepsilon_2(\omega)$ occurs in the lower photon energy. This suggests that the dielectric loss will be maximum at that peak energy. Furthermore, at ~30 eV the value of $\varepsilon_2(\omega)$ of both the compounds reduces to zero, which suggests that these compounds will be transparent above this energy. Despite the nearly identical shapes of the dielectric spectra of $X$Ir$_3$ compounds along [100] and [001] directions, there is some anisotropy over the whole energy/frequency range.

*(f) Energy loss function*

The energy loss function, $L(\omega)$ of a material provides valuable information about the electronic structure, plasmon resonance, absorption coefficient and reflectivity of the material [87,88]. This parameter also defines how much energy is lost when a fast electron traverses through the material. Furthermore, the plasma frequency ($\omega_p$) of a material is associated with the peak value of $L(\omega)$. The $L(\omega)$ spectra of $X$Ir$_3$ compounds are presented in Figures 9(f) and 10(f) along [100] and [001] polarization directions. The peak values of $L(\omega)$ of LaIr$_3$ compound in both the polarization directions are found at 32.8 eV. On the other hand, the peak values of $L(\omega)$ of ThIr$_3$ compound along [100] and [001] polarization directions are found at 32.3 and 31.8 eV, respectively. Thus, for ThIr$_3$ compound the value of $\omega_p$ is larger in [100] direction than [001] direction which indicates small amount of optical anisotropy in the compound. It can also be concluded from the Figures 9 and 10 that $\omega_p$ coincide with the sudden change in the absorption and reflectivity spectra of $X$Ir$_3$ compounds. Therefore, both the compounds will show optical properties similar to an insulator and will become transparent to incident light at energies above the $\omega_p$. Moreover, it is observed from the plots (see Figures 9 and 10) that in the lower photon energy (0 to 10 eV) the $L(\omega)$ spectra have no peaks. This is caused by the high value of $\varepsilon_2(\omega)$ in that energy range [89].

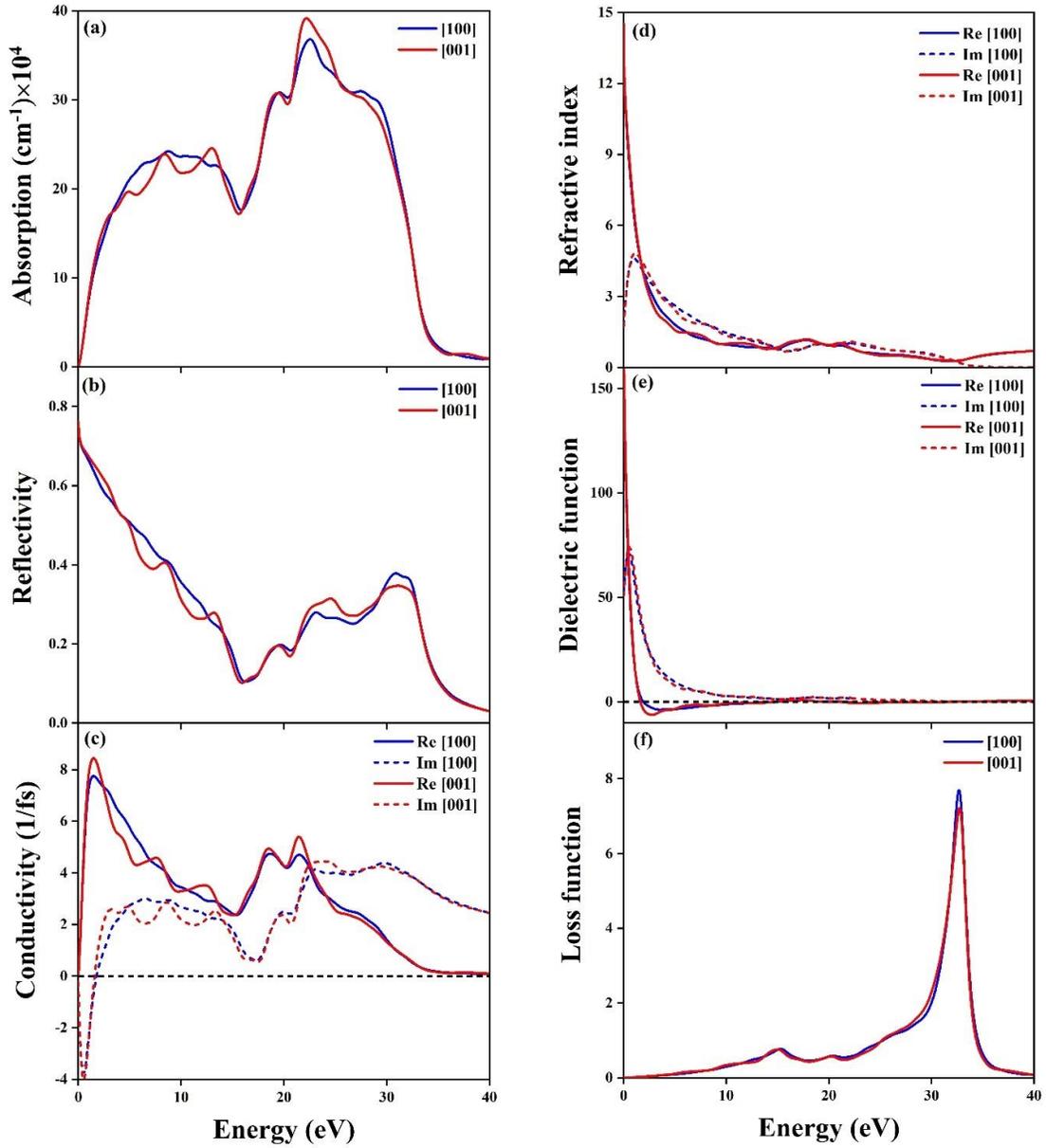

**Figure 9.** The computed (a) absorption coefficient, (b) reflectivity, (c) conductivity, (d) refractive index, (e) dielectric function, and (f) loss function of LaIr$_3$ along [100] and [001] polarization directions as a function of energy.

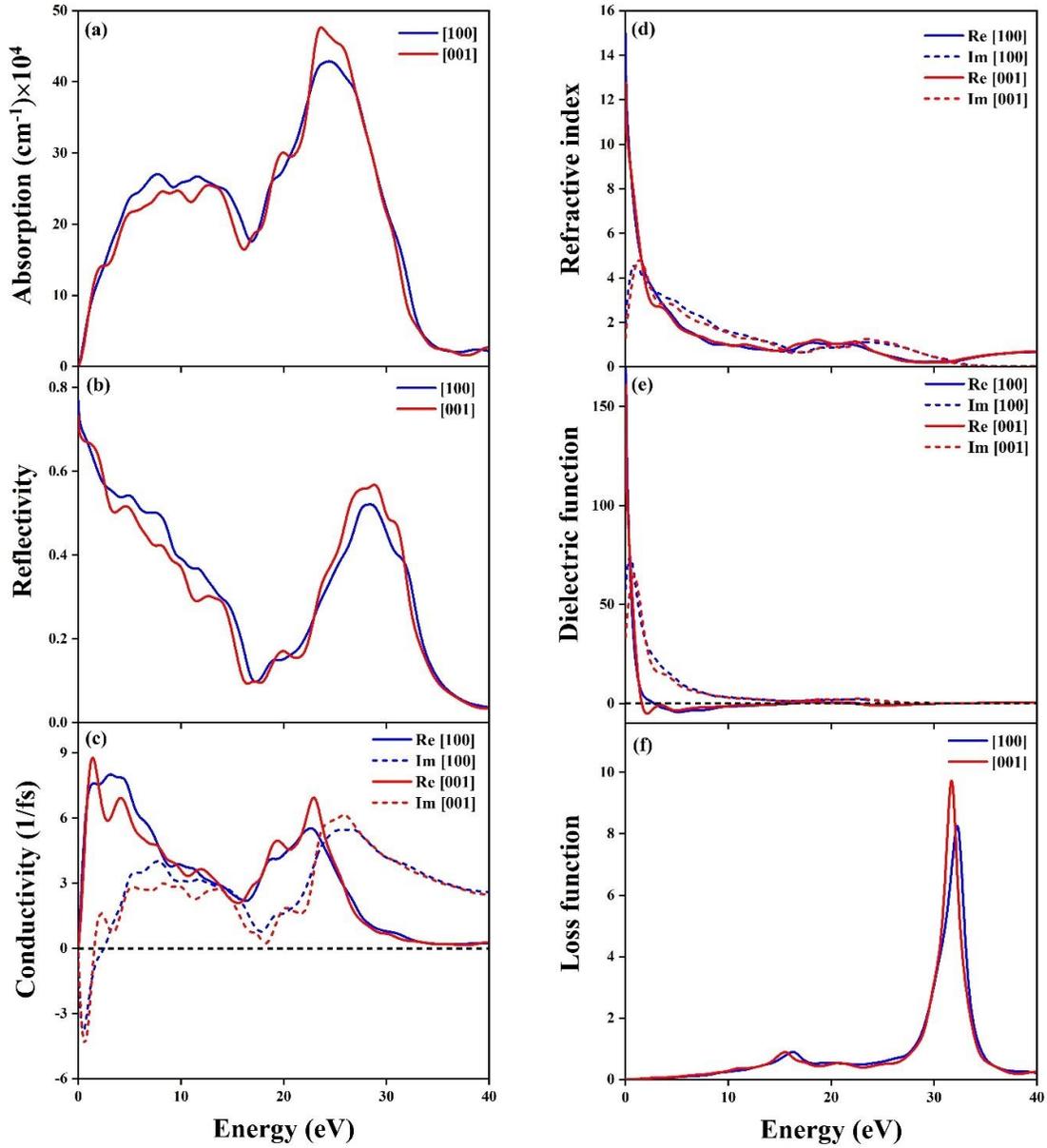

**Figure 10.** The computed (a) absorption coefficient, (b) reflectivity, (c) conductivity, (d) refractive index, (e) dielectric function, and (f) loss function of ThIr$_3$ along [100] and [001] polarization directions as a function of energy.

### 3.9. Superconducting state properties

Superconducting materials possess some very special electrical and magnetic properties which make them useful for a variety of technological applications. The superconducting state of a material is largely influenced by the electron-phonon coupling constant ($\lambda_{e-p}$) of the material. In this section, we have employed the well-known McMillan equation [90] to estimate the superconducting transition temperature ($T_c$) of the $X$Ir$_3$ ($X$ = La, Th) compounds.

$$T_c = \frac{\theta_D}{1.45} exp\left[\frac{-1.04(1 + l_{e\text{-}p})}{l_{e\text{-}p} - \mu^*(1 + 0.62 l_{e\text{-}p})}\right] \quad (60)$$

here, $\theta_D$ is the Debye temperature and $\mu^*$ refers to the repulsive screened Coulomb part (also known as the Coulomb pseudopotential) of the material. We have calculated the $\mu^*$ of $X$Ir$_3$ materials by employing the equation given below [91]:

$$\mu^* = \frac{0.26\, N(E_F)}{1 + N(E_F)} \quad (61)$$

where, $N(E_F)$ is the value of the TDOS of the compound at the Fermi level. The strength of the electron-phonon coupling of a material is described by the constant $\lambda_{e\text{-}p}$. In this study, we have calculated the $\lambda_{e\text{-}p}$ of $X$Ir$_3$ materials by using the following expression [92,93]:

$$l_{e\text{-}p} = \frac{\gamma_{exp}}{\gamma_{cal}} - 1 \quad (62)$$

In the above expression, $\gamma_{exp}$ is the experimental linear coefficient of the electronic specific heat and $\gamma_{cal}$ is its theoretical counterpart calculated from the TDOS at the Fermi level found from the band structure calculations. In this work, we have calculated $\gamma_{cal}$ of $X$Ir$_3$ materials by using the following equation [93–95]:

$$\gamma_{cal} = \frac{\pi^2 k_B^2 N(E_F)}{3} \quad (63)$$

The calculated values of superconducting parameters including the critical temperatures of $X$Ir$_3$ compounds are presented in Table 14. The Tc values obtained show reasonable agreement with the experimental ones [1,4].

For ThIr$_3$ compound, the value of $\lambda_{e\text{-}p}$ calculated from Eqn. (62) is in reasonable agreement with the value obtained from the Eliashberg spectral function. From the calculated values of $\lambda_{e\text{-}p}$, it can be suggested that both the compounds under study are moderately coupled electron-phonon superconductor. It is also noticed from the Table 14 that the computed values of $\mu^*$ of $X$Ir$_3$ compounds are much higher than those used in the previous studies [1,4]. In earlier works [1.4] the values of $\mu^*$ were selected ad-hoc without taking consideration of the material specific electronic features. Furthermore, it is found that the parameters ($\theta_D$, $\lambda_{e\text{-}p}$, and $\mu^*$) determining the superconducting critical temperatures of LaIr$_3$ and ThIr$_3$ are almost identical. It is instructive to note that the electron phonon coupling constant can be expressed as a product of two terms, $N(E_F)$ and $V_{e\text{-}ph}$ [96], where $V_{e\text{-}ph}$ is the electron-phonon interaction energy. Thus almost identical values of $\lambda_{e\text{-}p}$ of $X$Ir$_3$ ($X$ = La, Th) compounds imply that the electron-phonon interaction energy is stronger in ThIr$_3$, since $N(E_F)$ is found to be higher for LaIr$_3$.

**Table 14.** Superconducting state parameters of $X$Ir$_3$ ($X$ = La, Th) compounds along with the previously estimated values.

| Parameter | Unit | LaIr$_3$ | ThIr$_3$ | Ref. |
|---|---|---|---|---|
| $T_c$ | K | 4.91 | 5.01 | This work |
| | | 3.32 | 4.41 | [1,4] |
| $\mu^*$ | --- | 0.201 | 0.198 | This work |
| | | 0.15 | 0.13 | [1,4] |
| $\lambda_{e\text{-}p}$ | --- | 0.85 | 0.86 | This work |
| | | 0.57 | 0.74 | [1,4] |
| $\gamma_{cal}$ | mJ/(mol-K$^2$) | 8.17 | 7.57 | This work |
| $\gamma_{exp}$ | mJ/(mol-K$^2$) | 15.32 | 17.60 | [4,9] |

## 4. Conclusions

A large number of unexplored physical properties of $X$Ir$_3$ ($X$ = La, Th) compounds have been investigated in details. Some of the properties have been revisited and fair agreements with previous results are found. The compounds under study are mechanically stable, highly ductile, elastically anisotropic, and extremely machinable with moderate hardness. The computed Debye temperatures, minimum thermal conductivities are low. At the same time, the melting temperatures are high. High ductility, machinability, dry lubricity, and high fracture toughness indicates that the $X$Ir$_3$ ($X$ = La, Th) compounds have desirable traits for structural engineering applications. Low phonon thermal conductivity and high melting point imply that these to materials can be used for thermal insulation at high temperatures. The bonding nature has been investigated in details. Both the compounds show prominent ionic character with some covalent contributions.

The optical properties were investigated thoroughly for the first time. The optical parameters show weakly anisotropic features and reconfirm the metallic characters. The compounds are found to be efficient absorber of ultraviolet light. They are also good reflector ($R > 50\%$) of visible light that suggests that these materials can be used as efficient solar energy reflector. Very high values of low energy refractive index [$n(0) > 10$] further suggest that $X$Ir$_3$ ($X$ = La, Th) compounds possess features suitable for optoelectronic device applications. Nearly same plasma frequency for both the compounds indicates that carrier effective mass and concentrations are almost identical.

Highly dispersive bands are seen near the $\Gamma$-point in the band profiles of the compounds which implies low effective mass and high mobility of the charge carriers. It is found from the DOS profiles of $X$Ir$_3$ compounds that the Ir-$5d$ states largely contribute to the TDOS near the Fermi level. This indicates that the metallic conductivity in the compounds is governed mainly by the $5d$ states of Ir atom.

The calculated values of superconducting transition temperatures of $X$Ir$_3$ compounds are in close agreement with the previously estimated experimental values. Interestingly, the Fermi level of LaIr$_3$ is located at a rapidly changing part of the TDOS profile. Therefore, any shift in the Fermi energy due to applied pressure or suitable atomic substitution is expected to change the $N(E_F)$ significantly. Such modifications should significantly change the normal state charge transport properties, superconducting transition temperature, and superconducting condensation energy [97].

To conclude, a large number of physical features of the $X$Ir$_3$ ($X$ = La, Th) compounds have been investigated theoretically in this article. Most of the studied physical features are novel and can be used as references in further research.

**Acknowledgments**

Authors are grateful to the Department of Physics, Chittagong University of Engineering & Technology (CUET), Chattogram-4349, Bangladesh, for providing the computing facilities for this work. This work was also carried out with the aid of a grant (grant number: 21-378 RG/PHYS/AS_G-FR3240319526) from UNESCO-TWAS and the Swedish International Development Cooperation Agency (SIDA). The views expressed herein do not necessarily represent those of UNESCO-TWAS, SIDA, or its Board of Governors.

**Author Contributions**

**Md. Sajidul Islam:** Methodology, Formal analysis, Writing − original draft; **Razu Ahmed:** Methodology, Formal analysis; **M. A. Ali:** Formal analysis, Writing − original draft, Validation; **M. M. Hossain:** Formal analysis, Writing review & editing, Validation; **M. M. Uddin:** Writing review & editing, Validation; **S. H. Naqib:** Conceptualization, Formal analysis, Writing review & editing, Validation, Supervision.

**Declaration of interest**

The authors declare that they have no known competing financial interests or personal relationships that could have appeared to influence the work reported in this paper.

**Data availability**

The data sets generated and/or analyzed in this study are available from the corresponding author on reasonable request.

**List of References**